\documentclass[fleqn,usenatbib]{mnras}

\usepackage{amsmath}
\usepackage{graphicx}               
\usepackage[T1]{fontenc}
\hypersetup{linkcolor=blue}          
\usepackage{newtxtext,newtxmath}
\usepackage{enumitem}
\usepackage{multirow}
\usepackage{longtable}
\usepackage{color}
\usepackage{ulem}

\newcommand{\aref}[1]{\hyperref[#1]{Appendix~\ref*{#1}}}

\def\equationautorefname~#1\null{equation~(#1)\null}

\usepackage{scalerel,tikz}
\usetikzlibrary{svg.path}
\definecolor{orcidlogocol}{HTML}{A6CE39}
\tikzset{orcidlogo/.pic={
 \fill[orcidlogocol] svg{M256,128c0,70.7-57.3,128-128,128C57.3,256,0,198.7,0,128C0,57.3,57.3,0,128,0C198.7,0,256,57.3,256,128z};
 \fill[white] svg{M86.3,186.2H70.9V79.1h15.4v48.4V186.2z}
 svg{M108.9,79.1h41.6c39.6,0,57,28.3,57,53.6c0,27.5-21.5,53.6-56.8,53.6h-41.8V79.1z M124.3,172.4h24.5c34.9,0,42.9-26.5,42.9-39.7c0-21.5-13.7-39.7-43.7-39.7h-23.7V172.4z}
 svg{M88.7,56.8c0,5.5-4.5,10.1-10.1,10.1c-5.6,0-10.1-4.6-10.1-10.1c0-5.6,4.5-10.1,10.1-10.1C84.2,46.7,88.7,51.3,88.7,56.8z};
}}
\newcommand\orcidicon[1]{\href{https://orcid.org/#1}{\mbox{\scalerel*{
\begin{tikzpicture}[yscale=-1,transform shape]
\pic{orcidlogo};
\end{tikzpicture}
}{|}}}}

\title[Stellar dynamical modelling of 10K galaxies]{MaNGA DynPop -- I. Quality-assessed stellar dynamical modelling from integral-field spectroscopy of 10K nearby galaxies: a catalogue of masses, mass-to-light ratios, density profiles, and dark matter}

\author[K. Zhu et al.]{
Kai Zhu\orcidicon{0000-0002-2583-2669}$^{1,2,3}$\thanks{E-mail: \url{kaizhu@nao.cas.cn}},
Shengdong Lu\orcidicon{0000-0002-6726-9499}$^{4}$\thanks{E-mail: \url{lushengdong93@icloud.com}},
Michele Cappellari\orcidicon{0000-0002-1283-8420}$^{5}$,
Ran Li\orcidicon{0000-0003-3899-0612}$^{1,2,3}$\thanks{E-mail: \url{ranl@bao.ac.cn}},
Shude Mao\orcidicon{0000-0001-8317-2788}$^{4,1}$,
Liang Gao$^{1,2,3,6}$
\\
$^{1}$National Astronomical Observatories, Chinese Academy of Sciences, 20A Datun Road, Chaoyang District, Beijing 100101, China\\
$^{2}$Institute for Frontiers in Astronomy and Astrophysics, Beijing Normal University, Beijing 102206, China\\
$^{3}$School of Astronomy and Space Science, University of Chinese Academy of Sciences, Beijing 100049, China\\
$^{4}$Department of Astronomy, Tsinghua University, Beijing 100084, China\\
$^{5}$Sub-department of Astrophysics, Department of Physics, University of Oxford, Denys Wilkinson Building, Keble Road, Oxford, OX1 3RH, UK\\
$^{6}$Institute for Computational Cosmology, Department of Physics, University of Durham, South Road, Durham, DH1 3LE, UK\\
}

\date{Accepted 2023 April 18. Received 2023 April 18; in original form 2022 December 11}

\pubyear{2023}

\begin{document}
\label{firstpage}
\pagerange{\pageref{firstpage}--\pageref{lastpage}}
\maketitle

\begin{abstract}
This is the first paper in our series on the combined analysis of the Dynamics and stellar Population (DynPop) for the MaNGA survey in the final SDSS Data Release 17 (DR17). Here we present a catalogue of dynamically-determined quantities for over 10000 nearby galaxies based on integral-field stellar kinematics from the MaNGA survey. The dynamical properties are extracted using the axisymmetric Jeans Anisotropic Modelling (JAM) method, which was previously shown to be the most accurate for this kind of study. We assess systematic uncertainties using eight dynamical models with different assumptions. We use two orientations of the velocity ellipsoid: either cylindrically-aligned JAM$_{\rm cyl}$ or spherically-aligned JAM$_{\rm sph}$. We also make four assumptions for the models' dark vs. luminous matter distributions: (1) mass-follows-light, (2) free NFW dark halo, (3) cosmologically-constrained NFW halo, (4) generalized NFW dark halo, i.e. with free inner slope. In this catalogue, we provide the quantities related to the mass distributions (e.g. the density slopes and enclosed mass within a sphere of a given radius for total mass, stellar mass, and dark matter mass components). We also provide the complete models which can be used to compute the full luminous and mass distribution of each galaxy. Additionally, we visually assess the qualities of the models to help with model selections. We estimate the observed scatter in the measured quantities which decreases as expected with improvements in quality. For the best data quality, we find a remarkable consistency of measured quantities between different models, highlighting the robustness of the results.
\end{abstract}




\section{Introduction}
In the current Lambda cold dark matter cosmological paradigm, the mass budget in the Universe is dominated by dark matter (DM) existing in the form of dark matter halos. At the centre of the dark halos, the baryonic matter condenses and forms stars, giving birth to the galaxies we observe \citep{White1978,White1991}. The distributions of baryonic and dark matter play a crucial role in shaping the galaxies and thus are important for understanding galaxy formation and evolution mechanisms. However, due to the invisibility of dark matter, the observed galactic quantities are derived almost exclusively from observations of the baryonic matter, including gas and stars. A common approach to constrain the mass distribution of dark matter is based on the gravitational effect of dark matter on the motions of luminous tracers, i.e. stellar kinematics.

Since the many early studies revealing the existence of dark matter \citep[e.g.][]{Zwicky1933,Zwicky1937,Zwicky2009,Oort1940,Rubin1970,Bosma1979,Rubin1980,Bosma1981,Rubin1982,Rubin1985,Rubin1988}, galaxy rotation curves have been heavily used to probe the gravitational potential within which galaxies reside, and further constrain the mass distributions of galaxies. The rotation curve is usually derived from long-slit spectroscopy and only provides one-dimensional information on stellar kinematics. With the advent of Integral Field Unit (IFU) galaxy surveys, e.g. SAURON \citep{deZeeuw2002}, $\rm ATLAS^{3D}$ \citep{Cappellari2011}, CALIFA \citep{Sanchez2012}, SAMI \citep{Bryant2015}, MaNGA \citep{Bundy2015}, large samples of galaxies with spatially resolved stellar kinematics are observed. The two-dimensional kinematic information can be combined with dynamical modelling methods to investigate the mass distributions of galaxies, which put constraints on the formation and evolution of galaxies.

Current popular dynamical modelling methods include action-angle distribution function based methods \citep{Binney2010,Bovy2014}, orbit-based Schwarzschild methods \citep{Schwarzschild1979,Hafner2000,Gebhardt2001,Cappellari2006,vandenBosch2008,Long2018,ZhuL2018NatAs,ZhuL2020,ZhuL2022,Neureiter2021}, particle-based Made-to-Measure methods \citep{Syer1996,deLorenzi2007,Long2010,Hunt2014,Zhuling2014} and moment-based Jeans Anisotropic Modelling (JAM) method \citep{Cappellari2008,Cappellari2020}. Each method has its own advantage and can be applied under different conditions. The orbit-based and particle-based methods are more flexible to be applied to galaxies with triaxial shapes and give more detailed orbital information. 

However, the price to pay for this flexibility is an increased degeneracy of the resulting solutions. The degeneracy can be understood from simple dimensional arguments: the MaNGA data provide us the velocity moments at every position on the sky. This is a two-dimensional quantity, which cannot be expected to be sufficient to constrain both the three-dimensional distribution function of the stars and the two or three-dimensional gravitational potential \citep[see discussion in][sec.~3.4]{Cappellari2016ARA&A}. This implies that we need to make some assumptions to try to uncover trends in the data.

The JAM models make more restrictive assumptions but these were shown to capture remarkably well the high signal-to-noise ($S/N$) stellar kinematics of large samples of real galaxies \citep[see review by][]{Cappellari2016ARA&A}. 
More importantly, tests of JAM using high-resolution N-body simulations \citep{Lablanche2012} and cosmological hydrodynamical simulations \citep{Li2016} have confirmed its high accuracy and negligible bias in reproducing the total mass profiles.

More recently, JAM was compared in detail against the Schwarzschild method using samples of both observed galaxies, with circular velocities from CO gas \citep{Leung2018}, and numerical simulations \citep{Jin2019} respectively. For the comparison with CO gas circular velocities in real galaxies, considering the radial range of 0.8--1.6 $R_{\rm e}$, where the gas kinematics is well resolved and the rotation velocities are more accurately determined, the mean ratio for 54 galaxies between the errors of the Schwarzschild and of the JAM models is $\langle\sigma_{\rm SCH}/\sigma_{\rm JAM}\rangle\approx1.7$ \citep[fig.8 and table 4]{Leung2018}. Similarly, for the recovery of enclosed masses inside a sphere of radius $R_{\rm e}$ from simulations, considering all 45 model fits \citep[fig.~4]{Jin2019}, the 68\% percentile (1$\sigma$ error) of the absolute deviations between the model results and the true values is a factor 1.6 smaller for JAM than for Schwarzschild. In both cases, for both observations and simulations, there is no evidence for systematic biases between the true and recovered values. This higher accuracy of JAM justifies our choice of using it for this study.

JAM method has been successfully applied to many studies based on IFU data. For example, \citet{Cappellari2012Nature,Cappellari2013a,Cappellari2013b} construct detailed mass models, consisting of stellar and dark matter, for 260 early-type galaxies (ETGs) from $\rm ATLAS^{3D}$ survey \citep{Cappellari2011}, using JAM method. \citet{Cappellari2012Nature} study the stellar initial mass function (IMF) of galaxies by comparing the stellar mass-to-light ratios derived from JAM with those from stellar population synthesis (SPS), reporting a strong variation of stellar IMF normalization with the velocity dispersion within an effective radius $\sigma_{\rm e}$ for ETGs. This correlation between galaxy IMF and velocity dispersion of ETGs is then studied and confirmed in late-type galaxies \citep{Li2017}, galaxies in a denser environment \citep{Shetty2020}, and brightest cluster galaxies (BCGs; \citealt{Loubser2020,Loubser2021}), indicating a universal variation of stellar initial mass functions in various galaxies. Moreover, \citet{Cappellari2013a,Cappellari2013b}, \citet{Scott2015}, and \citet{Li2018a} make use of the JAM methods to study the scaling relations, such as fundamental planes (FPs), mass planes (MPs), mass-size planes, and so on. \citet{Cappellari2015} study the total mass-density slopes ($\gamma_{_{\rm T}}$) of 14 fast-rotating ETGs and report their nearly-isothermal total mass profiles, consistent with the values obtained from strong gravitational lensing analysis \citep{Auger2010}. \citet{Poci2017} and \citet{Li2019} enlarge the sample to  258 ETGs from $\rm ATLAS^{3D}$ and $\sim 2000$ galaxies (including ETGs and spirals) from MaNGA DR14 \citep{Abolfathi2018}, respectively, making it possible to analyse the correlations between galaxy total density slope and other galaxy properties, e.g. velocity dispersion, halo mass, stellar mass density, and so on. These studies together show the strong ability of JAM in analysing the structure and dynamics of galaxies, and the power in understanding the galaxy evolution.

With the final data release of the MaNGA project (SDSS DR17; \citealt{Abdurro'uf2022}), an unprecedentedly large sample (10K) of nearby galaxies with abundant information on the kinematics and stellar population properties is available. In this series of papers, we aim to build a complete library of dynamical models for these galaxies, including mass distributions (for both stellar and dark matter), dark matter fractions, stellar mass-to-light ratios, inclinations, and velocity asymmetry properties, using JAM with different assumptions. With these information, we will also investigate the scaling relations, galaxy mass density slopes, as well as initial mass functions of the complete MaNGA sample. Compared to previous studies (e.g. \citealt{Cappellari2012Nature,Cappellari2013a,Cappellari2013b,Cappellari2015,Scott2015,Li2017,Li2018a}; \citealt{Li2019,Shetty2020}), our sample size will be enlarged by at least a factor of five and reach 10K galaxies for the first time. The large sample span a wide range of galactic properties, e.g. stellar mass, morphology, and central velocity dispersion, making it a statistically significant sample to explore the correlations in great detail. 

In this paper (Paper I), we construct detailed JAM models for these 10000 nearby galaxies and provide their mass distributions of both stellar and dark matter components. The structure of this paper is as follows. In \autoref{sec:data}, we briefly introduce the MaNGA survey and the data used in this work. In \autoref{sec:method}, we describe the methods (i.e. JAM method and the mass models applied in this work) we used to derive our quantities. In \autoref{sec:cat}, \autoref{sec:quality_assessment}, and \autoref{sec:uncertainty}, we present the dynamical modelling results and investigate their robustness and systematic uncertainties. Finally, we summarize the results in \autoref{sec:sum}. Throughout the paper, we assume a flat Universe with $\Omega_{\rm m} = 0.307$ and $H_0 = 67.7\,\mathrm{km\cdot s^{-1}\cdot Mpc^{-1}}$ \citep{Planck2016}. The quantities in our catalogue can be converted to the standard flat cosmology with $\Omega_{\rm m}=0.3$ and $H_0=70\,\mathrm{km\cdot s^{-1}\cdot Mpc^{-1}}$, with very good accuracy, by multiplying enclosed masses by $K=67.7/70$, multiplying densities (in the unit of $\rm M_{\odot}\,pc^{-3}$) by $K^{-2}$, and multiplying mass-to-light ratios by $1/K$.

\section{Data}
\label{sec:data}
\subsection{The MaNGA survey}
The Sloan Digital Sky Survey-IV (SDSS-IV) Mapping Nearby Galaxies at Apache Point Observatory \citep[MaNGA;][]{Bundy2015} is an IFU survey which aims at obtaining spectral measurements across the face of $\sim$ 10K nearby galaxies. Using the tightly-packed arrays of optical fibers that feed into the BOSS spectrographs \citep{Smee2013,Drory2015} on the Sloan 2.5m telescope \citep{Gunn2006} at Apache Point Observatory, MaNGA provides the spatially resolved spectra that cover a radial range out to 1.5 effective radii ($R_{\rm e}$) for the Primary+ sample ($\sim 2/3$ of total sample) and out to 2.5 $R_{\rm e}$ for the Secondary sample ($\sim 1/3$ of total sample) at higher redshift \citep{Law2015,Wake2017}. The spaxel size of MaNGA is $0.5\arcsec$ and the average g-band Point Spread Function (PSF) FWHM (full width at half maximum) throughout the survey is about $2.54\arcsec$ \citep{Law2016}.

The spectra provided by MaNGA span a wavelength range of $3600-10300\,\Angstrom$ at a spectral resolution of $\sigma = 72\, \rm km/s$ \citep{Law2016}. Raw observational data are spectrophotometrically calibrated \citep{Yan2016} and processed by the Data Reduction Pipeline (DRP; \citealt{Law2016}) to produce three-dimensional data cubes.

\subsection{Stellar kinematics and imaging}
Higher-level products such as stellar kinematics, nebular emission-line properties, and spectral indices of the galaxies are produced by the Data Analysis Pipeline (DAP; \citealt{Belfiore2019,Westfall2019}). DAP derives the kinematic information from the IFU spectra of galaxies by fitting absorption lines using the \textsc{ppxf} software \citep{Cappellari2004,Cappellari2017,Cappellari2022} with the combination of a subset of the MILES \citep{Sanchez-Blazquez2006,Falcon-Barroso2011} stellar library, MILES-HC. Before fitting, the spectra are Voronoi binned \citep{Cappellari2003} to $\rm S/N = 10$ to ensure that the derived stellar velocity dispersions are reliable. The stellar velocity dispersion presented in DAP is a combination of $\sigma_*$, the intrinsic velocity dispersion of stars, and  $\sigma_{\rm diff}$, which is the quadrature difference between the instrumental dispersion of the galaxy template and the MaNGA data \citep{Westfall2019}. The velocity dispersion of the galaxy is given by $\sigma_*^2=\sigma_{\rm obs}^2-\sigma_{\rm diff}^2$. Note that the line-spread-function (LSF) of MaNGA survey has been improved significantly since MPL10 \citep{law2021}, the velocity dispersion measurements of this paper will be different from those used in \citet{Li2018a} especially for the dispersions far below instrumental resolution, which will lead to differences in dynamical modelling results.

In total, we have 10735 DAP outputs from SDSS DR17, within which the targets of ancillary programs (the Coma, IC342, M31, and globular clusters) should be excluded, resulting in 10296 datacubes for galaxy observations. Among the 10296 observations, we flag 151 datacubes that have been identified as critical-quality or unusual-quality by DRP, and the other 10145 datacubes are high-quality observations corresponding to 10010 unique galaxies and 135 repeat observations. We analyse 10296 datacubes in this work since the DRP quality flag is purposely conservative \citep{Westfall2019}, but users can use the DRP quality flag (see \aref{sec:appendix_catalogue}) to ignore the critical-quality or unusual-quality galaxies as a more conservative approach. We do not apply any other selection criteria, such that the sample enables a wide range of statistical analyses on galaxy properties as designed and can be corrected to a volume-limited sample \citep{Wake2017}. The final sample has a nearly flat stellar mass distribution in the range of $10^9-6\times10^{11} \,\rm M_{\odot}$ \citep{Wake2017} and has a median redshift of $z\sim0.03$.

In this work, we adopt the SDSS $r$-band \citep{Stoughton2002} PSF FWHM values from the catalogue\footnote{\url{https://www.sdss.org/dr17/manga/manga-data/data-access/}} provided by the MaNGA collaboration to account for the beam smearing effect on modelled stellar kinematics when comparing with the observed one \citep{Cappellari2008}. In addition, we make use of the $r$-band images and the corresponding PSF FWHM values\footnote{\url{https://www.sdss.org/dr12/imaging/images/}} of these $\sim 10000$ MaNGA galaxies from the SDSS data release 12 \citep{SDSSDR12} to obtain their surface brightness.

\section{Method}
\label{sec:method}
\subsection{Jeans Anisotropic Modelling}\label{sec:JAM}

In this section, we briefly introduce the mathematical basics of Jeans Anisotropic Modelling (JAM; \citealt{Cappellari2008,Cappellari2020}). For a steady-state axisymmetric stellar system, the Jeans equations in cylindrical coordinates, ($R, z, \phi$), are written as \citep[eq.~4–29a,c]{Binney1987}:
\begin{subequations}\label{eq:jeans_axi}
\begin{align}
    \frac{\nu\overline{v_{R}^2}-\nu\overline{v_{\phi}^2}}{R}+\frac{\partial(\nu\overline{v_{R}^2})}{\partial R}+\frac{\partial(\nu\overline{v_Rv_z})}{\partial z} &= -\nu\frac{\partial\Phi}{\partial R},\\
    \frac{\nu\overline{v_Rv_z}}{R}+\frac{\partial(\nu\overline{v_z^2})}{\partial z}+\frac{\partial(\nu\overline{v_Rv_z})}{\partial R} &= -\nu\frac{\partial \Phi}{\partial z},
\end{align}
\end{subequations}
where $\Phi$ is the gravitational potential, $\nu$ is the number density of the tracer population from which one measures the kinematics, and the notation related to the stellar distribution function $f$ is defined as
\begin{equation}
    \nu \overline{v_k v_j}\equiv\int v_k v_j f \ \mathrm{d}^3\mathbf{v}.
\end{equation}
This equation is quite general, as it {\em only} assumes axial symmetry and stead-state of the stellar system, but it has more unknowns than equations and cannot provide a unique solution.

To solve these equations, one has to make assumptions on the orientation of the velocity ellipsoid. \citet{Cappellari2008} assumes that the velocity ellipsoid is aligned with the cylindrical coordinates ($\overline{v_R v_z}=0$) and the velocity anisotropy in the meridional plane is constant\footnote{As will become clear later, the anisotropy is generally assumed constant for every component of the Multi-Gaussian Expansion, but does not have to be spatially constant for the whole galaxy.}, quantified as:
\begin{equation}
\label{eq:betaz}
    \beta_z \equiv 1-\frac{\overline{v_z^2}}{\overline{v_R^2}} = 1-\frac{\sigma_z^2}{\sigma_R^2} = 1-\frac{1}{b}.
\end{equation}
With these assumptions, \autoref{eq:jeans_axi} reduces to the following equations \citep[eq.~8,9]{Cappellari2008}
\begin{subequations}\label{eq:JAMcyl}
\begin{align}
	\frac{b\,\nu\overline{v_z^2}-\nu\overline{v_\phi^2}}{R}
	+ \frac{\partial(b\,\nu\overline{v_z^2})}{\partial R}
	&=  -\nu\frac{\partial\Phi}{\partial R}\\
	\frac{\partial(\nu\overline{v_z^2})}{\partial z}
	&=  -\nu\frac{\partial\Phi}{\partial z},
\end{align}
\end{subequations}
which have a unique solution for $\overline{v_z^2}$ and $\overline{v_\phi^2}$ under the boundary condition $\nu\overline{v_z^2}=0$ as $z\rightarrow\infty$.

The other extreme assumption to solve the Jeans equations is made by \citet{Cappellari2020}, where the velocity ellipsoid is set to be aligned with the spherical polar coordinate system ($r, \theta, \phi$) and the velocity anisotropy is defined as: 
\begin{equation}
\label{eq:betar}
    \beta_r \equiv 1-\frac{\overline{v_{\theta}^2}}{\overline{v_{r}^2}}=1-\frac{\sigma_{\theta}^2}{\sigma_{r}^2}.
\end{equation}
Under these assumptions, rewriting \autoref{eq:jeans_axi} in spherical coordinates and assuming a spherically-aligned velocity ellipsoid, the Jeans equations reduce to the following equations \citep[eq.~7]{Cappellari2020}
\begin{subequations}\label{eq:JAMsph}
	\begin{align}
		\frac{\partial(\nu\overline{v_r^2})}{\partial r} + \frac{(1+\beta_r)\,\nu\overline{v_r^2} - \nu\overline{v_\phi^2}}{r} & = -\nu\frac{\partial \Phi}{\partial r}\label{eq:jeans_beta_r}\\
		(1-\beta_r)\frac{\partial(\nu\overline{v_r^2})}{\partial \theta}  
		+ \frac{(1-\beta_r)\,\nu\overline{v_r^2} - \nu\overline{v_\phi^2}}{\tan\theta} & = -\nu\frac{\partial \Phi}{\partial \theta}\label{eq:jeans_beta_th},
	\end{align}
\end{subequations}
which have a unique solution for $\overline{v_r^2}$ and $\overline{v_\phi^2}$ under the boundary condition  $\nu\overline{v_r^2}=0$ as $r\rightarrow\infty$.

For a given gravitational potential $\Phi$, tracer density distribution $\nu$, and velocity anisotropy ($\beta_z$ or $\beta_r$), the intrinsic second velocity moments can be obtained by solving \autoref{eq:JAMcyl} or \autoref{eq:JAMsph}. The projected second velocity moments, $\overline{v_{\rm los}^2}$, can be then derived by integrating the intrinsic second velocity moments along the line-of-sight direction. By comparing the calculated line-of-sight second velocity moments and the observed root-mean-square velocity, $\rm V_{\rm rms}$ ($\equiv\sqrt{V^2+\sigma^2}$, where $\rm V$ and $\rm \sigma$ correspond to the stellar velocity and stellar velocity dispersion), we are able to constrain the model parameters (i.e. gravitational potential, velocity anisotropy, etc). We refer the readers to \citet{Cappellari2008,Cappellari2020} for more details of the JAM method, which is a formalism to efficiently solve \autoref{eq:JAMcyl} and \autoref{eq:JAMsph} using the Multi-Gaussian Expansion (MGE; \citealt{Emsellem1994,Cappellari2002}) as parametrization for both the tracer population and the total density. As mentioned by \citet{Cappellari2020}, different velocity ellipsoid assumptions may be appropriate for different types of galaxies. In this work, we adopt both assumptions to construct JAM models for the complete MaNGA sample such that the readers can check the robustness of derived quantities using the consistency of the two methods. We name the JAM  method with cylindrically-aligned velocity ellipsoid as JAM$_{\rm cyl}$ and the one with spherically-aligned velocity ellipsoid as JAM$_{\rm sph}$.
 
In practice, we calculate the predicted line-of-sight second velocity moments using the \textsc{python} version of JAM \textsc{JamPy}\footnote{Verison 6.3.3, available from \url{https://pypi.org/project/jampy/}}. The input total mass distribution $\Phi$ and tracer density $\nu$ of \textsc{JamPy} will be parameterized with MGE models. We assume a spatially constant velocity anisotropy to simplify the models, although \textsc{JamPy} allows for spatial variation of anisotropy by assigning different anisotropy values to the Gaussians of the input MGE model. 

\subsection{Multi-Gaussian Expansion}\label{sec:mge}
To obtain the tracer density distribution $\nu$, we perform Multi-Gaussian Expansion (MGE; \citealt{Emsellem1994,Cappellari2002}) fitting to the SDSS $r$-band images of the galaxies. The MGE formalism of surface brightness can be written as:
\begin{equation}
\label{eq:mge}
    \Sigma\left(x^{\prime}, y^{\prime}\right)=\sum_{k=1}^{N} \frac{L_{k}}{2 \pi \sigma_{k}^{2} q_{k}^{\prime}} \exp \left[-\frac{1}{2 \sigma_{k}^{2}}\left(x^{\prime 2}+\frac{y^{\prime 2}}{q_{k}^{\prime 2}}\right)\right],
\end{equation}
where $L_k$, $\sigma_k$, and $q^{\prime}_{k}$ are the total luminosity, dispersion along the major axis, and the axial ratio of $k$-th Gaussian component. To obtain the intrinsic luminosity density distribution from the deprojection of surface brightness, the galaxies are assumed to be oblate axisymmetric following \citet{Monnet1992}. Then, the 2D MGE can be deprojected to 3D space and the formalism in cylindrical coordinates can be written as \citep[eq.~13]{Cappellari2008}:
\begin{equation}
    v(R, z)=\sum_{k=1}^{N} \frac{L_{k}}{\left(\sqrt{2 \pi} \sigma_{k}\right)^{3} q_{k}} \exp \left[-\frac{1}{2 \sigma_{k}^{2}}\left(R^{2}+\frac{z^{2}}{q_{k}^{2}}\right)\right],
\end{equation}
where $L_k$ and $\sigma_k$ remain the same as the 2D MGE, while $q_k$ is the 3D intrinsic axial ratio of the $k$-th Gaussian component, written as:
\begin{equation}
    q_k = \frac{\sqrt{q^{\prime 2}_{k}-\cos^2 i}}{\sin i}.
\end{equation}
If we define the desired minimum intrinsic axial ratio of our deprojected MGE as $q_{\rm min}={\rm min}(q_k)$, we can calculate the inclination ($i=90 ^\circ$ being edge-on) of the MGE by inverting the previous equation as:
\begin{equation}
	\label{eq:inc}
	\tan^2 i = \frac{1-q^{\prime2}_{\rm min}}{q^{\prime2}_{\rm min}-q_{\rm min}^2},
\end{equation}
where $q^{\prime}_{\rm min}={\rm min}(q^{\prime}_k)$ is the minimum axial ratio of the projected 2D Gaussian components.

Some of the galaxies in our sample are weakly triaxial or prolate. However, the triaxial galaxies are generally close to spherical in their central parts \citep{Emsellem2011,Cappellari2016ARA&A} and are expected to be well approximated by nearly spherical $\rm JAM_{sph}$ models. Prolate galaxies are extremely rare \citep{Li2018b,Krajnovic2018} and unlikely to play any significant role in our trends.

\begin{figure*}
    \centering
    \includegraphics[width=\textwidth]{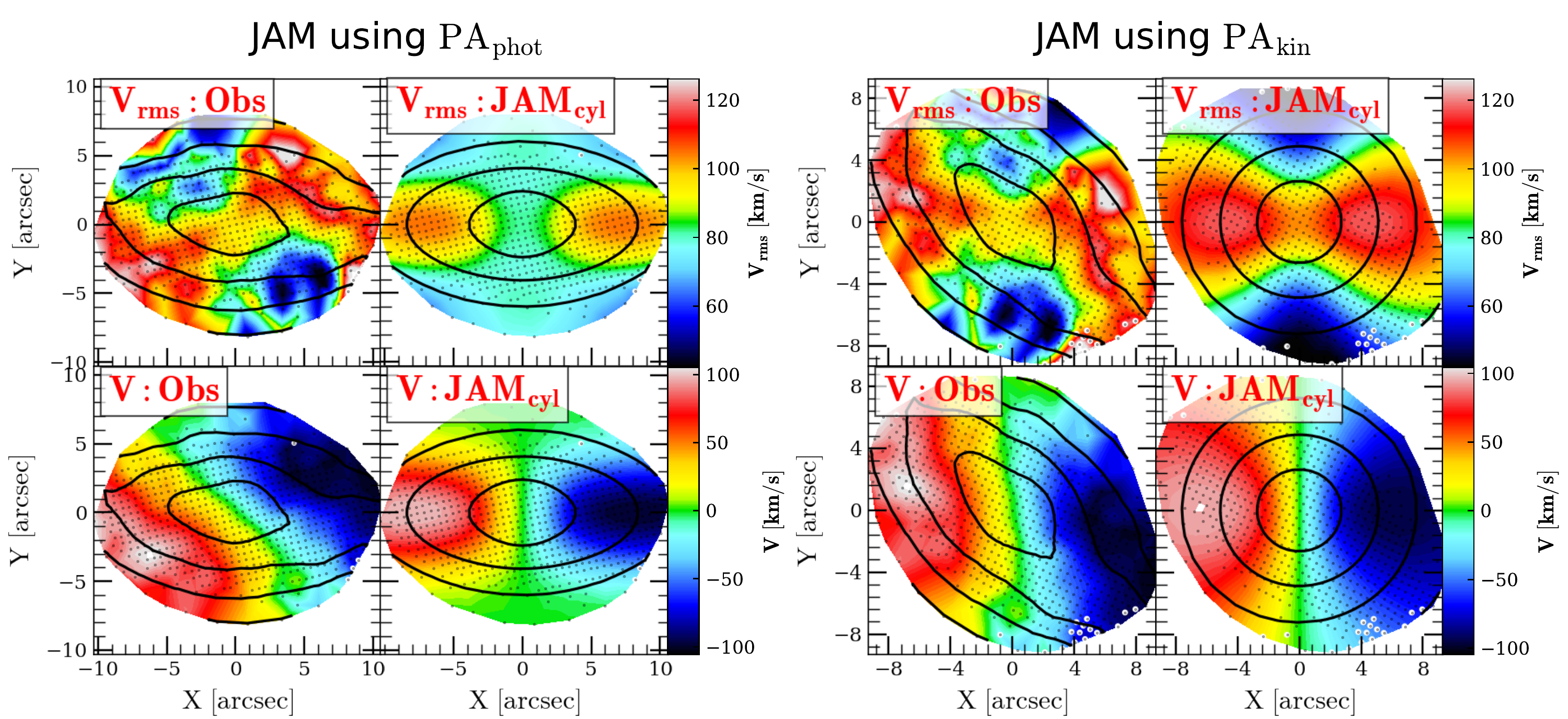}
    \caption{An Example of the comparison between aligning models with the photometric or kinematic position angles (PA) for a barred galaxy (MaNGA ID: 1-155463). The left sub-figure is the fitting result of JAM$_{\rm cyl}$ + MFL model (see \autoref{sec:method} for definition) by assuming the galaxy major axis is aligned with the bar, which here defines the photometric PA. The right sub-figure shows the model results by aligning the JAM model with the kinematic PA. In each sub-figure, the observed second velocity moment $V_{\rm rms}$ and the light-of-sight velocity $\rm V$ are shown in the left panels, and the corresponding modelled maps are in the right panels. All the kinematic maps are oriented in the way that the major (photometric or kinematic) axis is oriented along the x-direction. The black contours are the observed (left panels) and modelled (right panels) surface brightness contours in steps of 1 mag, respectively. The black dots are the centroids of the Voronoi bins from which the maps were linearly interpolated.}
    \label{fig:kinpa_mgepa}
\end{figure*}

In this work, we make use of the \textsc{python} package \textsc{mgefit}\footnote{Version 5.0.13, from  \url{https://pypi.org/project/mgefit/}} which implements the method by \citet{Cappellari2002} to do the MGE fitting. Before fitting, we first go through the images of the galaxies visually and mask the foreground stars or  galaxies if necessary. Then, the galaxy center and the ellipticity are derived by the \textsc{find\_galaxy} routine in the package. We note here that the position angles (PA) used in the MGE fitting are normally the photometric ones (i.e. derived from the galaxy $r$-band image). However, we find that many of the MaNGA galaxies have different kinematic PA (calculated from their line-of-sight velocity map) and photometric PA, although the PA discrepancies of most galaxies are small. The reason for these differences is generally due to bars \citep[e.g.][]{Krajnovic2011}. As JAM assumes an axisymmetric shape, the modelled velocity field is aligned with the modelled photometry in the MGE formalism. In the presence of bars, the stellar kinematics gives a better estimate for the position angle of the underlying stellar disk than the photometry. Thus, in this work, we use the kinematic PA derived by \textsc{pafit}\footnote{Version 2.0.7, from \url{https://pypi.org/project/pafit/}} package, which implements the method by \citet[appendix~C]{Krajnovic2006}, in the MGE fitting process, except for those galaxies with highly disturbed velocity field (visually checked). We show in \autoref{fig:kinpa_mgepa} the comparison of JAM results based on the photometric PA and kinematic PA for the misaligned galaxy. As can be seen, for such a kinematic-photometric misaligned galaxy, a kinematic PA-based MGE results in a better root-mean-square velocity ($V_{\rm rms}$) recovery. We also confirm visually that the kinematic position angle also helps to avoid the bars or other non-axisymmetric features in the automated procedure of MGE fitting and reflects the underlying mass distribution of the whole galaxy.

\begin{figure*}
    \centering
    \includegraphics[width=\textwidth]{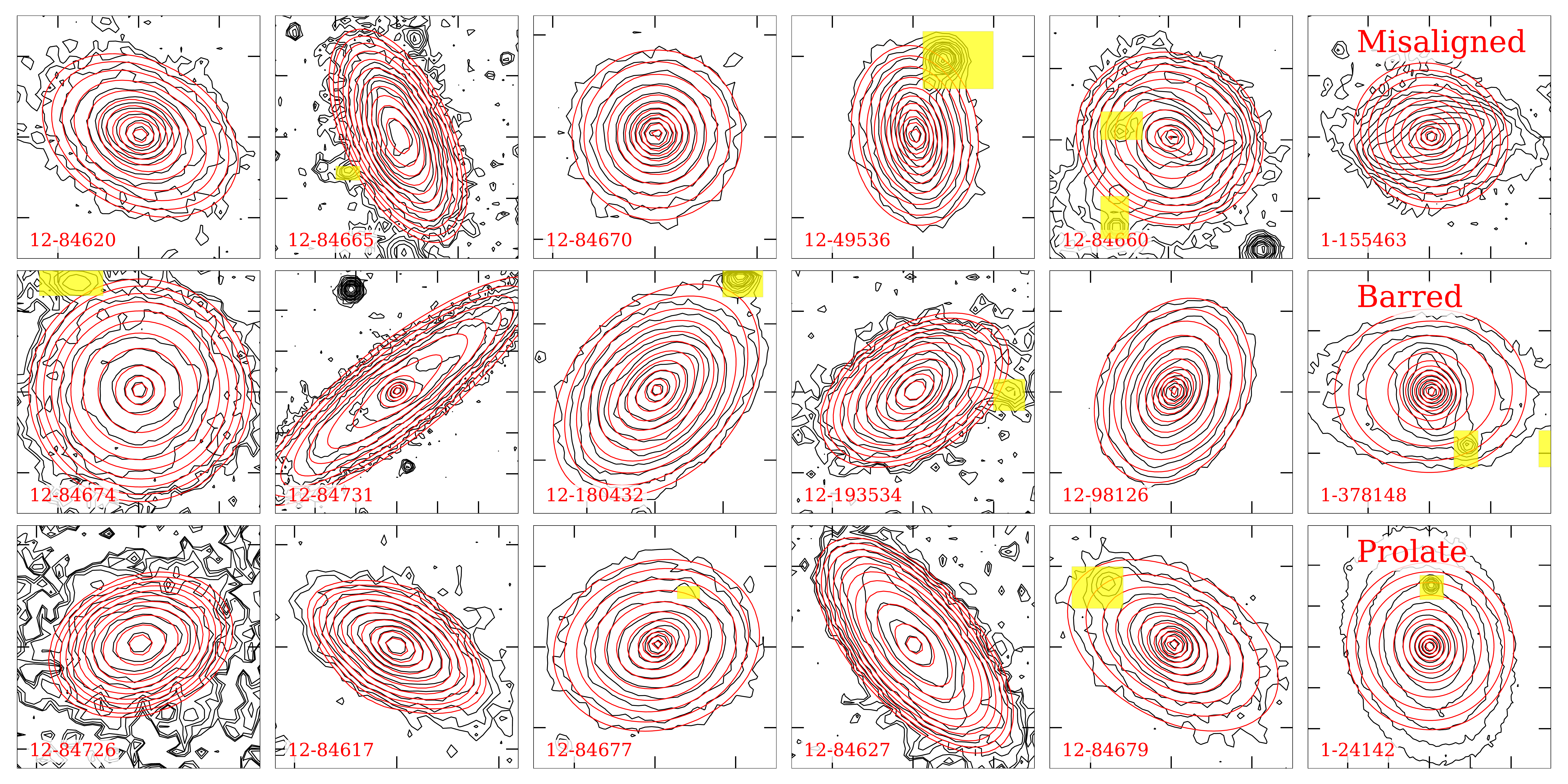}
    \caption{SDSS $r$-band photometry of example galaxies (black curves) with MGE model contours (red curves) over-plotted. MaNGA ID is presented in each panel. The first five columns: randomly selected galaxies with similar photometric position angles and kinematic position angles. The rightmost column shows a photometric-kinematic misaligned galaxy, a galaxy with a strong bar, and a prolate galaxy from top to bottom. In each panel, the yellow shaded regions are the masked regions, which are excluded in MGE fitting. Each tick in all panels is 10 arcsec. In all the fittings, the kinematic position angle is used (see \autoref{sec:mge}).}
    \label{fig:mge_example}
\end{figure*}

Moreover, we make use of the \textsc{mge\_fit\_sectors\_regularized} routine instead of \textsc{mge\_fit\_sectors} to fit the MGE formalism of galaxy photometry. This implements the method of \citet[sec.~3.2.1]{Scott2013} to reduce the influence of bars on the model results. \textsc{mge\_fit\_sectors\_regularized} here provides a narrower range of axial ratios for different Gaussian components of MGE. The MGEs so fitted artificially remove the flattest components while keeping the fitting solution only slightly changed (within the agreement of measurement uncertainties). In \autoref{fig:mge_example}, we show several example MGE models for different kinds of galaxies (e.g. normal galaxies, kinematic-photometric misaligned galaxies, barred galaxies, and prolate galaxies).

Before performing JAM, the MGE surface brightness is multiplied by a factor of $(1+z)^3$, with $z$ the galaxy redshift, to account for both the bolometric cosmological dimming effect and the change of bandwidth in the AB magnitude system. Considering that most of the MaNGA galaxies locate at low redshift, we do not apply K-correction \citep{Hogg2002}. Throughout this work, we adopt the $r$-band absolute solar magnitude as  $\rm M_{\odot,r}=4.65$ in AB system \citep{Willmer2018}.

\subsection{Mass models and parameter design}
\label{sec:model_design}
The gravitational potential, $\Phi$, is determined by the total mass distribution, consisting of an extended matter distribution (stellar and dark matter) and a supermassive black hole at the center of the galaxy. The mass of black hole is estimated using the $\rm M_{BH}-\sigma_{c}$ relation \citep{McConnell2011Nature}, where $\sigma_{c}$ is computed as mean stellar velocity dispersion within an aperture of radius equal to the FWHM of the MaNGA PSF. Below, we describe the mass models adopted in this work.

\subsubsection{Model A: Mass-follows-light JAM model}
\label{sec:model_A}
Mass-follows-light model (MFL, hereafter) is also known as the self-consistent model (e.g. \citealt{Cappellari2013a,Shetty2020}). In this model, we assume that both the distribution of the kinematic tracer population $\nu$ and the total mass of the galaxy follows the $r$-band luminosity distribution and thus can be obtained from the luminosity MGEs by multiplying a constant dynamical mass-to-light ratio, $M/L$. Therefore, there are three free parameters that are needed to fit in order to match the observed $V_{\rm rms}$: (1) the velocity dispersion ratio $\sigma_{z}/\sigma_{R}$ in JAM$_{\rm cyl}$ (or $\sigma_{\theta}/\sigma_{r}$ in JAM$_{\rm sph}$), (2) the minimum intrinsic axial ratio $q_{\rm min}$ which we use to parametrize the inclination, and (3) the dynamical mass-to-light ratio $M/L$. 

\subsubsection{Model B: JAM model with NFW dark halo}
This model allows for the decomposition between the stellar mass distribution and dark matter mass distribution. In this case, the kinematic tracer population $\nu$ is still deprojected from the observed stellar surface brightness. However, the mass distribution has extra freedom. It is described as a superposition of the deprojected stellar luminosity, scaled by a spatially constant stellar mass-to-light ratio $M_{\ast}/L$, plus a spherical  NFW \citep{Navarro1996} dark halo. Contrary to what is sometimes incorrectly stated, this kind of model is formally correct even in the presence of spatial $M/L$ gradients in the galaxy. In fact, with $M/L$ variations the tracer population $\nu$ is still best approximated by the surface brightness (not by the surface density) and the model would still return a formally correct estimate of the total density. What would be biased in this case is our interpretation of the total density decomposition in terms of luminous and dark components.
The NFW profile is written as
\begin{equation}
    \rho_{_{\rm DM}}(r)=\rho_s\left(\frac{r}{r_s}\right)^{-1}\left(\frac{1}{2}+\frac{1}{2}\frac{r}{r_s}\right)^{-2},
\end{equation}
where $r_s$ is the characteristic radius and $\rho_s$ is the characteristic density. To reduce the number of free parameters, we calculate the characteristic radius $r_s$ with scaling relations, instead of setting it to be a free parameter: at each step of parameters optimisation procedure, $r_s$ is calculated by the redshift-dependent stellar-to-halo mass relation \citep{Moster2013} and the redshift-dependent mass-concentration relation \citep{Dutton2014b}, with a given stellar mass (or stellar mass-to-light ratio). We fix the limits of the $r_s$ to be $5\times r_{\rm max, bin}<r_s<250$ kpc, with $r_{\rm max, bin}$ the largest radius of the Voronoi bins, to avoid unrealistic small/large $r_s$. The lower limit of $5\times r_{\rm max, bin}$ is approximately the median value of the $r_{\rm s, NSA}$ distribution, which are estimated using the Chabrier-IMF stellar mass and the scaling relations described above. The upper limit of 250 kpc is large enough to ensure that the halo profiles with $r_s>250$ kpc are nearly identical (at least within the kinematic data range of MaNGA) to the one with $r_s=250$ kpc. The choice of $r_s$ is not a critical assumption: the characteristic radius of halo is much larger (at a median ratio of 5) than the kinematic data range of MaNGA, indicating that the halo profile within observed region of MaNGA can be described as a simple power law density profile. Therefore, this model has four free parameters: (1) the minimum intrinsic axial ratio $q_{\rm min}$, (2) the anisotropy $\sigma_{z}/\sigma_{R}$ or $\sigma_{\theta}/\sigma_{r}$, (3) the stellar mass-to-light ratio $M_*/L$, and (4) the characteristic density of NFW halo $\rho_s$. 

\subsubsection{Model C: JAM model with fixed NFW dark halo}
The model is similar to Model B, but assumes an NFW dark halo profile without any free parameter. Both $\rho_s$ and $r_s$ are inferred from the scaling relations used in Model B, which means that the inclusion of dark halo is predicted by the simulations. Thus, there are only three free parameters of this model: (1) $q_{\rm min}$, (2) $\sigma_{z}/\sigma_{R}$ in JAM$_{\rm cyl}$ (or $\sigma_{\theta}/\sigma_{r}$ in JAM$_{\rm sph}$), and (3) $M_*/L$.

\subsubsection{Model D: JAM model with gNFW dark halo}
We also adopt a more general dark halo profile, the generalized NFW \citep[gNFW,][]{Wyithe2001} profile, in this work to allow for the baryonic effect on the inner density slope of dark halo \citep{Abadi2010,Duffy2010,Laporte2012}. The only difference between this model and Model B is the inner mass density slope within the characteristic radius of the dark halo. The gNFW profile can be written as:
\begin{equation}
\rho_{_{\rm DM}}(r) = \rho_s\left(\frac{r}{r_s}\right)^{\gamma}\left(\frac{1}{2}+\frac{1}{2}\frac{r}{r_s}\right)^{-\gamma-3},
\end{equation}
where $\gamma=-1$ results in an NFW profile, similar to Model B. The variation of $\gamma$ allows for more kinds of dark halos, including some special cases such as the ones with an inner core ($\gamma\approx 0$). We estimate the characteristic radius $r_s$ here the same as Model B, and thus there are five free parameters in this model: (1) $q_{\rm min}$, (2) $\sigma_{z}/\sigma_{R}$ in JAM$_{\rm cyl}$ (or $\sigma_{\theta}/\sigma_{r}$ in JAM$_{\rm sph}$), (3) $M_*/L$, (4) $\rho_s$, and (5) $\gamma$. 

For simplicity, we name the four JAM models (A-D) following their assumptions on the dark halos, i.e. MFL model = Model A, NFW model = Model B, fixed NFW model = Model C, and gNFW model = Model D.

\subsubsection{Parameter limits}\label{sec:para_bounds}
We note here that, with given $r_s$ and $\gamma$ (being $-1$ for an NFW dark halo), the characteristic density $\rho_s$ can be uniquely calculated from the stellar mass-to-light ratio and the dark matter fraction within a given aperture. Thus, we take the dark matter fraction within a sphere with its radius being the 2D projected half-light radius $R_{\rm e}$, $f_{\mathrm{DM}}(<R_{\rm e})$ as the free parameter in NFW and gNFW models, instead of $\rho_s$. In this way, we do not need to worry about the boundary of $\rho_s$, but simply set the range $0<f_{\mathrm{DM}}(<R_{\rm e})<1$. The 2D effective radius $R_{\rm e}$ here is computed using \textsc{mge\_half\_light\_isophote} software in \textsc{JamPy} package and then multiplied by 1.35 as described in \autoref{sec:size}.

\begin{figure*}
    \centering
    \includegraphics[width=2.0\columnwidth]{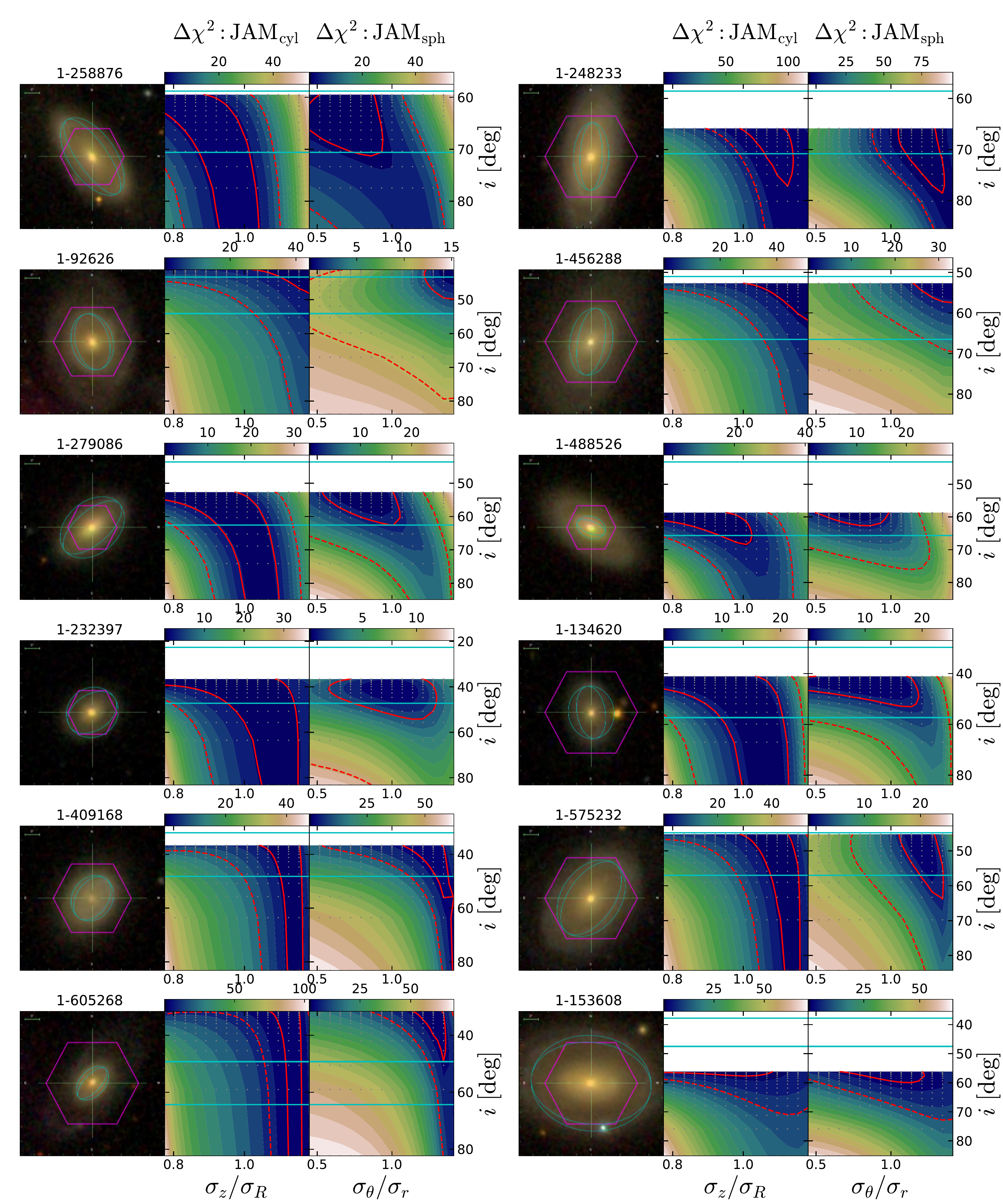}
    \caption{The $\Delta\chi^2$ ($\equiv \chi^2-\chi^2_{\rm min}$) distributions on the inclination-velocity anisotropy planes for 12 selected galaxies with dust rings (see \autoref{sec:model_design} for model design). $\chi^2$ values are the fitted residuals of MFL models with a given inclination-velocity anisotropy parameter pair; $\chi^2_{\rm min}$ is the minimum value of all the possible $\chi^2$. In each sub-figure, the left panel shows the image of the galaxy with the two ellipses indicating the upper and lower limit of apparent axial ratios. Results of JAM$_{\rm cyl}$ and JAM$_{\rm sph}$ are shown in the middle and right panels of each sub-figure, with the horizontal blue lines indicating the range of inclination angle determined from the dust geometry (see \autoref{sec:para_bounds} for details). The red solid and dashed contours represent 1$\sigma$ ($\Delta\chi^2=2.3$) and 3$\sigma$ ($\Delta\chi^2=11.8$) confidence levels of the fitting. The degeneracy between the velocity anisotropy and the inclination is strong for the JAM$_{\rm cyl}$, while the degeneracy is weaker for JAM$_{\rm sph}$, requiring more(less) restrictive boundaries of velocity anisotropy for JAM$_{\rm cyl}$ (JAM$_{\rm sph}$).}
    \label{fig:inc_beta_chi2}
\end{figure*}

In addition, the boundary of velocity anisotropy parameter is important for recovering the galaxy inclination (or equivalently the minimum intrinsic axial ratio $q_{\rm min}$). In previous studies with $\rm JAM_{cyl}$, the boundary of $\beta_z$ was constrained to be $\beta_z>0$ (equivalently, $\sigma_{z}/\sigma_{R}<1$) to break the inclination-anisotropy degeneracy \citep{Cappellari2008}. A similar criterion cannot be applied to the $\rm JAM_{sph}$ models and for this reason we explore if we can place limits on $\sigma_{\theta}/\sigma_r$, to ensure that the $\rm JAM_{sph}$ inferred inclination is consistent with the inclination determined from the geometry of dust disks in galaxies. For the MaNGA galaxies, we visually select 12 galaxies with dust rings. The dust rings are assumed to be circular and thus the observed ellipticity of the ring is thought to be only from the effect of the inclination. We fit ellipses to the extracted dust ring to obtain the upper and lower limit of the axial ratio and then calculate the range of galaxy inclination. More details about the procedure can be found in \aref{sec:appendix_dustring}. In \autoref{fig:inc_beta_chi2}, we show the contours of $\Delta\chi^2$ on the $(\sigma_z/\sigma_R,i)$, or $(\sigma_{\theta}/\sigma_r,i)$ plane. $\Delta\chi^2$ is defined as:
\begin{equation}
\Delta\chi^2 \equiv \chi^2-\chi^2_{\rm min},
\end{equation}
where $\chi^2$ is the fitted residual of the MFL model (both JAM$_{\rm cyl}$ and JAM$_{\rm sph}$) with a set of given $[\sigma_z/\sigma_R,i]$ (or $[\sigma_z/\sigma_R,i]$ in JAM$_{\rm sph}$); $\chi^2_{\rm min}$ is the minimum value of all the possible $\chi^2$. As can be seen, for JAM$_{\rm cyl}$, most of the galaxies show a strong degeneracy between $\sigma_z/\sigma_R$ and $i$ and only assuming $\sigma_z/\sigma_R<1$ can the JAM inferred inclination match the one derived from the dust ring, in agreement with \citet{Cappellari2008}. For JAM$_{\rm sph}$, however, the degeneracy between $\sigma_{\theta}/\sigma_r$ and $i$ is weaker. Therefore, there is no need to set a strict upper limit on $\sigma_{\theta}/\sigma_r$ because the best-fit model can always recover the galaxy inclination. In this work, we adopt a rather extreme upper limit $\sigma_{\theta}/\sigma_r<2$ ($\beta_r>-3$) that has never been observed in real galaxies. 

In the course of this project, we had to run the JAM models multiple times while trying to understand the differences between model results and optimize the reliability of our extracted parameters. We discovered cases where the JAM$_{\rm cyl}$ and JAM$_{\rm sph}$ models provided quite different total density slopes. Analysing these deviant cases, we realized this was due to the mass-anisotropy degeneracy \citep{Binney1982,Gerhard1993} allowing for unrealistic values of the anisotropy and making the JAM$_{\rm sph}$ return unreliable density profiles, due to modest quality data. Rather than simply exclude those galaxies as unreliable, we used our extensive knowledge of the anisotropy, accumulated from many detailed dynamical models of nearby galaxies, as a prior to reduce this degeneracy and extract as much useful information as we could from the data.

In particular, we now know that fast rotator galaxies satisfy an approximate upper limit in anisotropy \citep[e.g.][fig.~9]{Cappellari2016ARA&A}, which has recently been explained as due to a physical limit on high-anisotropy equilibrium solutions in flat galaxies \citep{Wang2021}. The limit has the form \citep[eq.~11]{Cappellari2016ARA&A}
\begin{equation}\label{eq:beta_eps}
	\beta\la0.7\times\varepsilon_{\rm intr},
\end{equation}
and appears valid for both cylindrical and spherical alignment of the velocity ellipsoid \citep{Wang2021}. 

Rewriting \autoref{eq:beta_eps}, the empirical lower boundary of velocity dispersion ratio is: 
\begin{equation}
\label{eq:delta}
    \mathcal{R}(q) =\sqrt{0.3+0.7q},\qquad
    \sigma_z/\sigma_R \ga \mathcal{R}(q),\qquad
    \sigma_{\theta}/\sigma_r \ga \mathcal{R}(q),
\end{equation}
where $q$ is the characteristic intrinsic axial ratio of the galaxy. These limits are derived for fast rotators galaxies, however they are also known to apply to slow rotators, which are generally close to spherical in the central parts and close to isotropic \citep{Gerhard2001,Gebhardt2003,Cappellari2007,Thomas2009}.
In conclusion, we restrict $\mathcal{R}(q)<\sigma_z/\sigma_R<1$ and $\mathcal{R}(q)<\sigma_{\theta}/\sigma_r<2$ in this work. The parameters and their boundaries of different JAM models are summarized in \autoref{tab:model}.

The characteristic intrinsic axial ratio $q$ is estimated as follows: deprojecting the observed MGE with the inclination angle derived from the initial fitting procedure using MFL models (\autoref{sec:fitting procedure}), then re-projecting the intrinsic MGE with $i=90^{\circ}$ and obtaining the ellipticity of the re-projected MGE, $\varepsilon_{\rm intr}$, from the \textsc{mge\_half\_light\_isophote} software in the \textsc{JamPy} package, finally the intrinsic axial ratio $q$ is calculated by $q\equiv1-\varepsilon_{\rm intr}$.

\begin{table*}
\centering
\caption{Parameter design, parameter boundary, the optimisation methods for 8 models (see \autoref{sec:JAM} and \autoref{sec:model_design}). $q^{\prime}_{\rm min}$ is the minimum observed axial ratio, $\mathcal{R}(q)$ is the empirical lower limit of velocity dispersion ratio determined from \autoref{eq:delta}.}
\label{tab:model}
    \begin{tabular}{|l|c|c|r|c|c|r|}
        \hline
        \multicolumn{1}{c|}{\multirow{2}*{Models}} &\multicolumn{5}{|c}{Free parameters}&\multirow{2}*{optimisation method} \\
        \cline{2-6}
        \multicolumn{1}{c|}{} & $q_{\rm min}$ & $\sigma_z/\sigma_R( \sigma_{\theta}/\sigma_r$) & $\lg M/L [{\rm M_{\odot}/L_{\odot}}]$ & $f_{\rm DM} (<R_{\rm e})$ & $\gamma$\\
        \hline
        $\rm (A1)\ JAM_{cyl}$ + MFL & [0.05, $q'_{\rm min}$] & $\sigma_z/\sigma_R$: [$\mathcal{R}(q)$, 1] & $\lg (M/L)_{\rm e}: [-2, 2]$ & - & - &\textsc{direct} + \textsc{least-squares}\\
        $\rm (A2)\ JAM_{sph}$ + MFL & [0.05, $q'_{\rm min}$] & $\sigma_{\theta}/\sigma_r$: [$\mathcal{R}(q)$, 2] & $\lg (M/L)_{\rm e}: [-2, 2]$ & - & - &\textsc{direct} + \textsc{least-squares}\\
        $\rm (B1)\ JAM_{cyl}$ + NFW & [0.05, $q'_{\rm min}$] & $\sigma_z/\sigma_R$: [$\mathcal{R}(q)$, 1] & $\lg M_*/L: [-2, 2]$ & [0, 1] & - &\textsc{direct} + \textsc{least-squares}\\
        $\rm (B2)\ JAM_{sph}$ + NFW & [0.05, $q'_{\rm min}$] & $\sigma_{\theta}/\sigma_r$: [$\mathcal{R}(q)$, 2] & $\lg M_*/L: [-2, 2]$ & [0, 1] & - &\textsc{direct} + \textsc{least-squares}\\
        $\rm (C1)\ JAM_{cyl}$ + fixed NFW & [0.05, $q'_{\rm min}$] & $\sigma_z/\sigma_R$: [$\mathcal{R}(q)$, 1] & $\lg M_*/L: [-2, 2]$ & - & - &\textsc{direct} + \textsc{least-squares}\\
        $\rm (C2)\ JAM_{sph}$ + fixed NFW & [0.05, $q'_{\rm min}$] & $\sigma_{\theta}/\sigma_r$: [$\mathcal{R}(q)$, 2] & $\lg M_*/L: [-2, 2]$ & - & - &\textsc{direct} + \textsc{least-squares}\\
        $\rm (D1)\ JAM_{cyl}$ + gNFW & [0.05, $q'_{\rm min}$] & $\sigma_z/\sigma_R$: [$\mathcal{R}(q)$, 1] & $\lg M_*/L: [-2, 2]$ & [0, 1] & [-1.6, 0] & Qual=0: \textsc{direct} + \textsc{least-squares} \\&&&&&& $\rm Qual\geqslant1$: \textsc{multinest}\\
        $\rm (D2)\ JAM_{sph}$ + gNFW & [0.05, $q'_{\rm min}$] & $\sigma_{\theta}/\sigma_r$: [$\mathcal{R}(q)$, 2] & $\lg M_*/L: [-2, 2]$ & [0, 1] & [-1.6, 0] & Qual=0: \textsc{direct} + \textsc{least-squares} \\&&&&&& $\rm Qual\geqslant1$: \textsc{multinest}\\
        \hline
    \end{tabular}
\end{table*}

\subsection{Parameter optimizations}
\subsubsection{optimization tools}
We use two different tools to find the best-fit parameters of the JAM models. The first is a least-squares fitting method. In this work, we make use of the \href{https://docs.scipy.org/doc/scipy/reference/generated/scipy.optimize.least_squares.html}{\texttt{scipy.optimize.least\_squares}} routine of the \textsc{python} software Scipy\footnote{\url{https://scipy.org}} \citep{Scipy2020}. To avoid the possibility that the fitted parameters are stuck in a local minimum, we further adopt the \href{https://docs.scipy.org/doc/scipy/reference/generated/scipy.optimize.direct.html}{\texttt{scipy.optimize.direct}} global optimisation algorithm \citep{Jones1993}, as originally implemented by \citet{Gablonsky2001}, to find a starting guess for the least-squares optimization.

To be able to assess the degeneracies and formal uncertainties in the model parameters from the posterior distribution, as well as to be more robust against the possibility of missing the global minimum, we also employ a Bayesian inference method. According to the Bayes theorem, the posterior probability distribution of the model  with a given set of parameters $\boldsymbol{p}$, given a set of data $\boldsymbol{d}$ is:
\begin{equation}
    P(\boldsymbol{p}|\boldsymbol{d}) \propto P(\boldsymbol{d}|\boldsymbol{p})\times P(\boldsymbol{p}),
\end{equation}
where $P(\boldsymbol{d}|\boldsymbol{p})$ is the likelihood of the data, given some model,  and $P(\boldsymbol{p})$ is the prior on our model. Assuming that the observational errors are Gaussian, the likelihood $P(\boldsymbol{d}|\boldsymbol{p})$ is proportional to $\exp{\left(-\frac{\chi^2}{2}\right)}$, where $\chi^2$ is defined as: 
\begin{equation}
    \chi^2 = \sum_{j}\left[\frac{\left({\overline{v^{2}_{\textrm{ los}}}}_{,j}\right)^{1/2}-V_{\textrm {rms},j}}{\varepsilon_{_{V_{\textrm{rms},j}}}}\right]^2,
\end{equation}
where $\overline{v^{2}_{\textrm{ los}}}_{,j}$ is the modelled second moment velocity of $j$-th Voronoi bin; $V_{\textrm{rms},j}$ and $\varepsilon_{_{V_{\textrm{rms},j}}}$ are the corresponding observed root-mean-square velocity and $1\sigma$ uncertainties (see \autoref{sec:fitting procedure} for more details on error estimation). The summation goes over all the Voronoi bins of MaNGA DAP outputs. In this work, we adopt the software \textsc{pymultinest}\footnote{Version 2.11, \url{https://pypi.org/project/pymultinest/}} \citep{Buchner2014}, a \textsc{python} version of the efficient and robust Bayesian tool \textsc{multinest} \citep{Feroz2009,Feroz2019}, which is a multimodal nested sampling algorithm, to carry out the Bayesian inference of the model parameters. The priors needed in the Bayesian inference are set to be flat within the given boundary as shown in \autoref{tab:model}. The best-fit model refers to the set of parameters that has the maximum likelihood (corresponding to maximum posterior probability in the case of flat priors). We adopt 500 active points and 0.8 sampling efficiency for \textsc{pymultinest} in this work.

\begin{figure}
    \centering
    \includegraphics[width=\columnwidth]{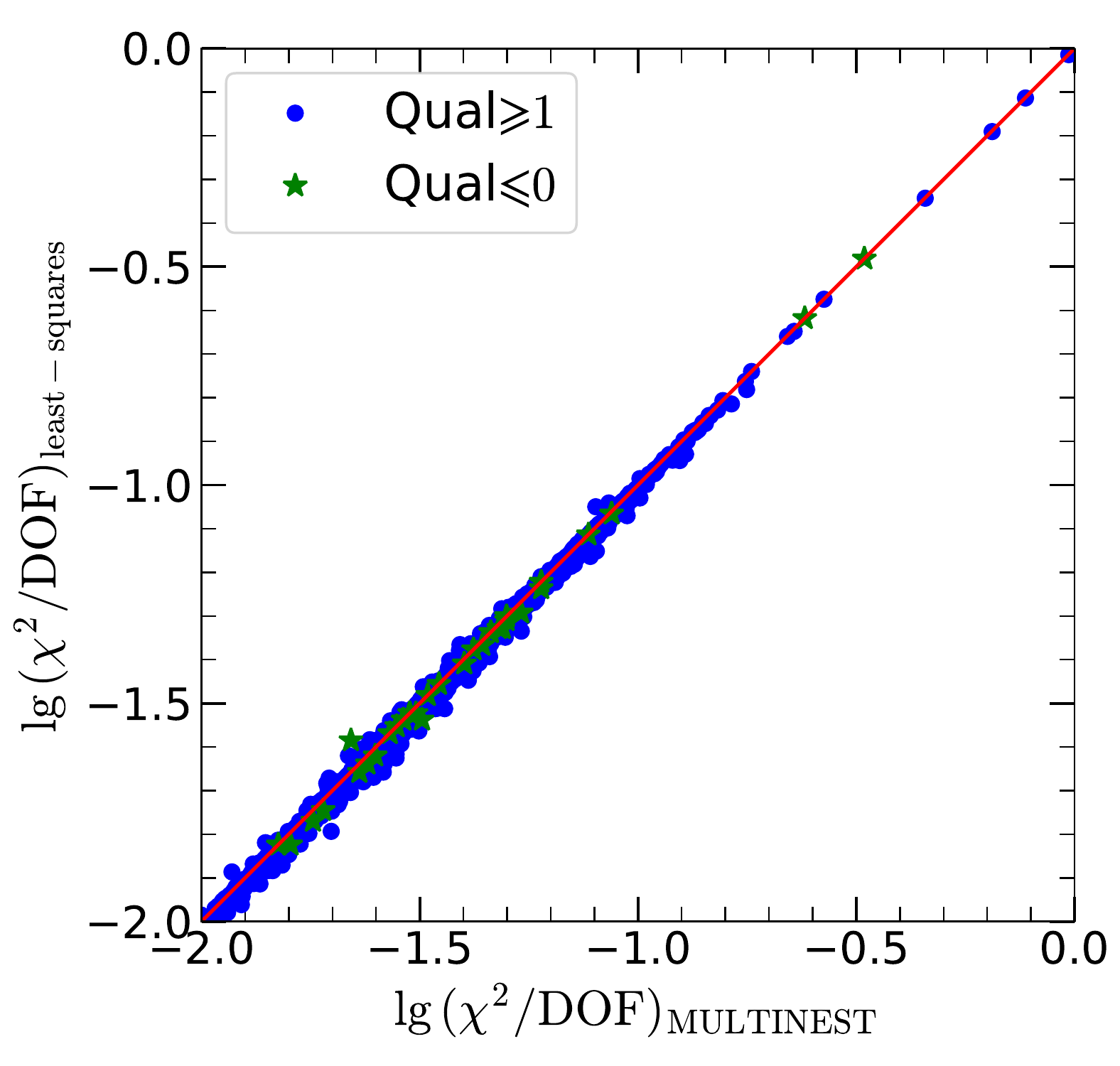}
    \caption{Comparisons of the $\rm \chi^2/DOF$ derived from \textsc{multinest} (X-axis) and from least-squares fitting (Y-axis). The JAM method with cylindrically-aligned velocity ellipsoid (i.e. JAM$_{\rm cyl}$, see \autoref{sec:JAM} for definition) is applied on the gNFW model (see \autoref{sec:model_design}) for 6065 $\rm Qual\geqslant 1$ galaxies (blue circles). We also randomly select 100 galaxies, among which are $\sim40$ $\rm Qual\leqslant 0$ galaxies (green stars), to investigate the convergence of optimisation methods for galaxies with bad modelling quality. $\rm Qual\geqslant 1$ and $\rm Qual\leqslant 0$ refer to the better modelling qualities and worse modelling qualities, respectively (see the definitions in \autoref{sec:quality}). The red solid line represents the $y=x$ relation.}
    \label{fig:lsq_mul_gNFW}
\end{figure}

In \autoref{fig:lsq_mul_gNFW}, we present the comparison between the $\rm \chi^2/DOF$ derived from the least-squares fitting and the Bayesian inference, adopting the JAM$_{\rm cyl}$ method and the gNFW mass model. As can be seen, the $\rm \chi^2/DOF$ are highly consistent between different optimisation methods for the $\rm Qual\geqslant1$ galaxies (see \autoref{sec:quality} for the definitions of $\rm Qual$). To investigate the convergence of optimisation methods for the galaxies with bad modelling quality, we randomly select 100 galaxies ($\sim40$ $\rm Qual\leqslant0$ galaxies) from the complete MaNGA sample and find a similar level of high convergence. In summary, we apply least-squares fitting on the complete sample for all models, but only apply the Bayesian inference on the $\rm Qual\geqslant1$ galaxies for the gNFW model to obtain the posterior distributions of parameters (both JAM$_{\rm cyl}$ and JAM$_{\rm sph}$).

\subsubsection{Fitting procedure}
\label{sec:fitting procedure}
\begin{figure}
    \centering
    \includegraphics[width=\columnwidth]{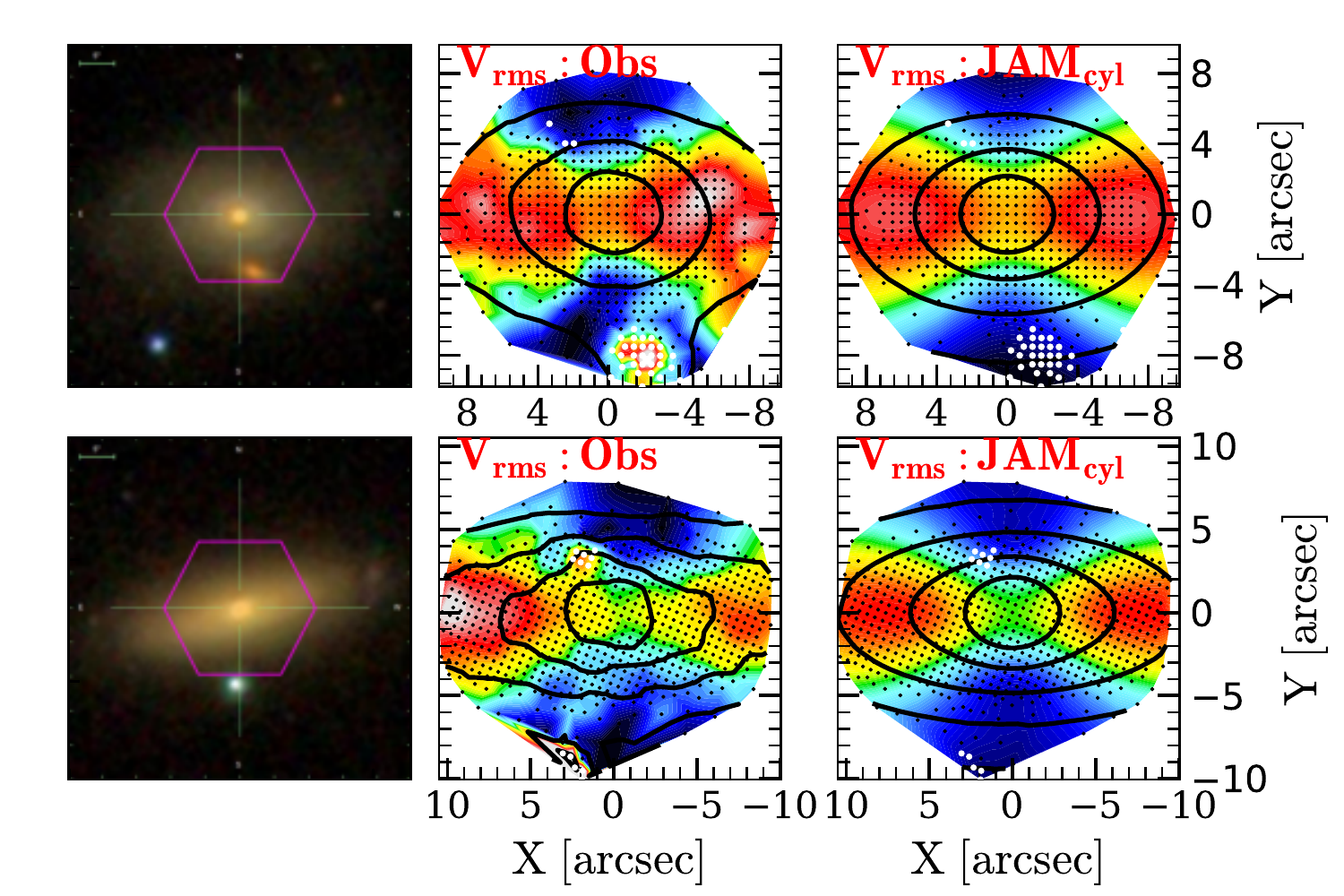}
    \caption{Examples of pixel clipping in JAM modelling. The white circles in the velocity fields are the clipped bins, corresponding to the spurious kinematics features caused by satellite galaxies or foreground stars, which are not fitted in JAM.}
    \label{fig:clip}
\end{figure}

The fitting procedure can be divided into two steps. In the first step, we perform an initial fitting with JAM$_{\rm cyl}$ on a MFL model (\autoref{tab:model}), which aims at removing some spurious kinematics features like stars or the problematic bins of the data and determining the kinematic errors. The first aim of this step is achieved by iteratively fitting the kinematics and clipping the bins with deviation beyond 3$\sigma$ confidence of the noise, until convergence. We compute the noise as a bi-weight estimate \citep{Hoaglin1983} from the differences of the data and model $V_{\rm rms}$. Then the remained bins will be adopted in the following processes of optimisation, including either the least-squares fitting or the Bayesian inference. Two examples of the clipped kinematic maps are shown in \autoref{fig:clip}. 

We also determine the kinematic errors $\varepsilon_{_{V_{\rm rms}}}$ from the initial fitting with JAM$_{\rm cyl}$ + MFL model in the first step. In order to prevent the dynamical models from being strongly affected by the inner high-$\rm S/N$ Voronoi bins, we do not use the true observed kinematic errors. Instead, we follow the practice in \citet{Mitzkus2017}, where the initial kinematic errors of $V_{\rm rms}$ are derived by error propagation as:  
\begin{equation}
    \varepsilon^{\rm init}_{_{V_{\rm rms}}} = \frac{1}{V_{\rm rms}}\sqrt{(V\ \varepsilon_V)^2 +(\sigma\varepsilon_{\sigma})^2},
\end{equation}
where the errors on velocity and velocity dispersion ($\varepsilon_V$ and $\varepsilon_{\sigma}$) are set to be $\rm \varepsilon_V = 5 \ km\ s^{-1}$ and $\rm \varepsilon_{\sigma}=0.05\sigma$, respectively. After the initial fitting process with $\varepsilon^{\rm init}_{_{V_{\rm rms}}}$, we compute the $\chi^2$ and scale the uncertainties using:
\begin{equation}
    \varepsilon_{_{V_{\rm rms}}} = \varepsilon^{\rm init}_{_{V_{\rm rms}}} \times \sqrt{\rm \chi^2/DOF} \times (2N_{\rm bins})^{1/4},
\end{equation}
where $\rm \chi^2/DOF$ is the reduced chi-square value of the initial fitting and $\rm N_{bins}$ is the number of remained kinematic bins after the initial fitting. The two scaling factors of $\sqrt{\rm \chi^2/DOF}$ and $(2N_{\rm bins})^{1/4}$  are used to first increase the kinematic errors to reach $\rm \chi^2/DOF=1$ and second to try to account, in an approximate way, for possible systematic uncertainties \citep{Mitzkus2017}. This approach is a Bayesian implementation of the idea that was suggested for confidence levels based on  $\chi^2$ by \citet{vandenBosch2009} and was since used on numerous papers based on large IFS datasets. 

In the second step, the clipped kinematic maps and the renewed kinematic errors $\varepsilon_{_{V_{\rm rms}}}$ are used in the final parameter estimation (both for the least-squares fitting and the Bayesian inference) to obtain the final results of the eight models listed in \autoref{tab:model}.

\section{The catalogue}
\label{sec:cat}

We create a catalogue containing the dynamical properties of MaNGA galaxies derived from JAM method. The complete list together with brief descriptions is shown in \aref{sec:appendix_catalogue}. We clarify the calculations of some of the derived quantities below.

\subsection{Size parameters}
\label{sec:size}
We provide three size parameters in this catalogue, namely $R_{\rm e}^{\rm maj}$, $R_{\rm e}$, and $r_{1/2}$. Here $R_{\rm e}^{\rm maj}$ is the major axis of the half-light elliptical isophote; $R_{\rm e}$ is the circularized effective radius, satisfying that the area of half-light ellipse $A=\pi R_{\rm e}^2$, and $r_{1/2}$ is the radius of the 3D sphere which encloses half of the total luminosity of the galaxy. For spherical galaxies $r_{1/2}\approx1.33R_{\rm e}$ for a wide range of profiles shape \citep{Ciotti1991}. However, for more general flattened galaxies the ratio between these two radii varies wildly with inclination and shape \citep[fig.~4]{Cappellari2013a}. The projected size parameters $R_{\rm e}$ and $R_{\rm e}^{\rm maj}$ are calculated from the MGE formalism of the galaxy $r$-band luminosity distribution, using the \textsc{mge\_half\_light\_isophote} software in \textsc{JamPy}. The $r_{1/2}$ is computed by linear interpolation on the deprojected luminosity profile, which is derived for a set of given radii using the \textsc{mge\_radial\_mass} software and adopting the best-fit inclination for each model (thus the $r_{1/2}$ values for different models are similar but not totally the same). In \citet{Cappellari2013a}, the 2D effective radii (i.e. $R_{\rm e}^{\rm maj}$ and $R_{\rm e}$), which are derived using the same photometric data (SDSS r-band imaging) and technique, are scaled by a factor of 1.35 to match the values determined from 2MASS \citep{Skrutskie2006} plus RC3 \citep{deVaucouleurs1991}. We adopt the same correction and all quantities related to the 2D effective radii in this catalogue always use the scaled radii.

\subsection{Mass and mass density slopes}
For all of the 8 models (4 mass models with both JAM$_{\rm cyl}$ and JAM$_{\rm sph}$), we provide total mass within given apertures and total mass slopes. For the 6 models allowing decomposition of dark matter and stellar mass, we also derived  mass and slopes for dark matter and stellar component separately. 

Enclosed masses (stellar, dark matter) are calculated analytically using the \textsc{mge\_radial\_mass} routine in \textsc{JamPy}, within 2 different spheres [i.e. with radii being $R_{\rm e}$ and $r_{1/2}$, respectively], denoted as $M_{\rm T}(<R_{\rm e})$ [or $M_{\ast}(<R_{\rm e})$ and $M_{\rm DM}(<R_{\rm e})$ for stellar and dark matter] and $M_{\rm T}(<r_{1/2})$ [or $M_{\ast}(<r_{1/2})$ and $M_{\rm DM}(<r_{1/2})$ for stellar and dark matter]. We also give the dynamical mass-to-light ratio $(M/L)_{\rm e}$ as the ratio between $M_{\rm T}(<R_{\rm e})$ and the $r-$band luminosity within a sphere of $R_{\rm e}$\footnote{Note that the so-calculated dynamical mass-to-light ratio coincides with the free parameter $M/L$ used in the MFL model.}.

We calculate two kinds of density slopes. One is the average logarithmic slope $\gamma_{_{\rm T}}$ \citep{Cappellari2015,Poci2017}, which can be written as 
\begin{equation}\label{eq:density2}
    \gamma_{_{\rm T}} = \frac{1}{\lg (R_{\rm out}/R_{\rm in})} \int_{R_{\rm in}}^{R_{\rm out}} \frac{\mathrm{d}\lg \rho_{_{\rm T}}}{\mathrm{d}\lg r} \mathrm{d}\lg r = \frac{\lg \rho_{_{\rm T}}(R_{\rm out})-\lg \rho_{_{\rm T}}(R_{\rm in})}{\lg R_{\rm out}-\lg R_{\rm in}},
\end{equation}
where $R_{\rm out}$ is set to be $R_{\rm e}$, and $R_{\rm in}$ is set to be the maximum between $0.1R_{\rm e}$ and the FWHM of MaNGA PSF.

\begin{figure}
    \centering
    \includegraphics[width=\columnwidth]{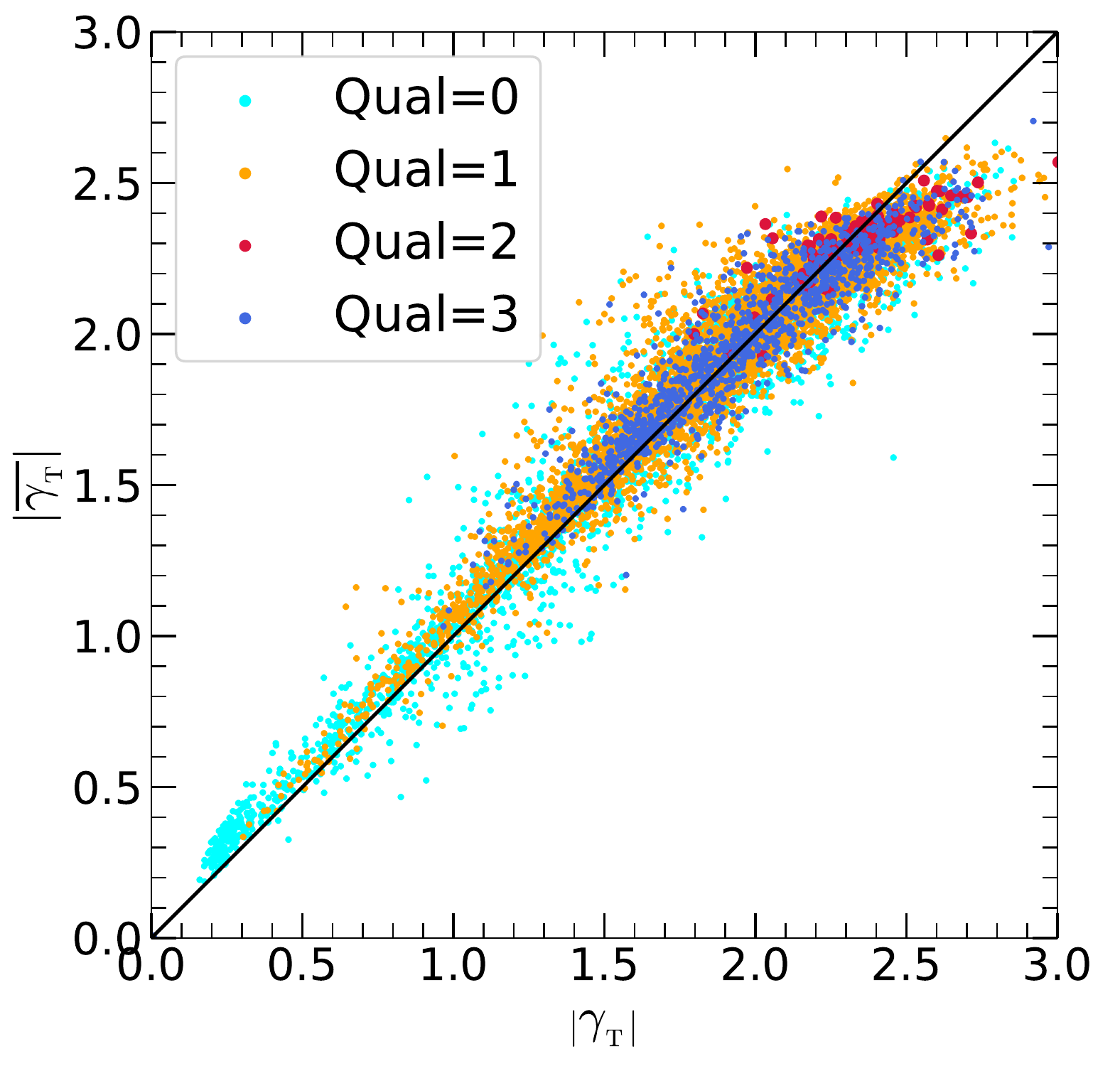}
    \caption{The comparison of the absolute values between two definitions of total density slopes: the average logarithmic slope $\gamma_{_{\rm T}}$ (x-axis) and the mass-weighted slope $\overline{\gamma_{_{\rm T}}}$ (y-axis). The symbols with different colours represent the galaxies with different modelling qualities (see \autoref{sec:quality} for the definitions). The black straight line is the $y=x$ relation.}
    \label{fig:cmp_slopes}
\end{figure}

The other one is the mass-weighted density slope $\overline{\gamma_{_{\rm T}}}$ \citep{Dutton2014a}, given by
\begin{equation}
    \overline{\gamma_{_{\rm T}}} \equiv \frac{1}{M_{\rm T}(<R_{\rm e})}\int_0^{R_{\rm e}}-\frac{\mathrm{d}\lg{\rho_{_{\rm T}}}}{\mathrm{d}\lg{r}} 4\pi r^2\rho_{_{\rm T}}(r)\mathrm{d}r = 3-\frac{4\pi R_{\rm e}^3\rho_{_{\rm T}}(R_{\rm e})}{M_{\rm T}(<R_{\rm e})}\,.
\end{equation}
The mass density slopes for the stellar ($\gamma_{_{\ast}}$, $\overline{\gamma_{_{\ast}}}$) and dark matter components ($\gamma_{_{\rm DM}}$, $\overline{\gamma_{_{\rm DM}}}$) can also be calculated similarly. In all cases, the radially averaged density is computed analytically from the axisymmetric MGE using the \textsc{mge\_radial\_density} function within \textsc{JamPy}. We provide slopes of both definitions in the catalogue, but only show the mass-weighted slopes in the analysis in the following sections. Note that the two definitions have opposite signs: the average logarithmic slope is negative, while the mass-weighted one is positive. A comparison between the two definitions of the total density slopes (the absolute values) is presented in \autoref{fig:cmp_slopes}.

\subsection{Effective velocity dispersion and stellar angular momentum}
\label{sec:sigma_lambda}
The effective second moments $\langle v_{\rm rms}^2\rangle_{\rm e}$ is defined as the luminosity-weighted second velocity moments within an elliptical aperture of area $A = \pi R_{\rm e}^2$, which reads:
\begin{equation}\label{eq:sigmae}
   \langle v_{\rm rms}^2\rangle_{\rm e} = \frac{\sum_k F_k (V_k^2+\sigma_k^2)}{\sum_k F_k},
\end{equation}
where $F_k$, $V_k$, and $\sigma_k$ are the flux, stellar velocity, and stellar velocity dispersion in the $k$-th spaxel, respectively. As $V$ and $\sigma$ are derived from the \textsc{ppxf} fit to the Voronoi binned spectra, we replicate the binned values for each spaxel belonging to each Voronoi bin. The effective velocity dispersion $\sigma_{\rm e}$, which is measured from a single fit to the stacked spectra within half-light ellipse, is found to be close to $\langle v_{\rm rms}^2\rangle_{\rm e}$ within the random errors \citep{Cappellari2013a}. The proxy for the stellar angular momentum $\lambda_{R_{\rm e}}$ is defined as \citep{Emsellem2007}:
\begin{equation}\label{eq:lambdaRe}
    \lambda_{R_{\rm e}} = \frac{\sum_k F_k R_k|V_k|}{\sum_k F_k R_k \sqrt{V_k^2+\sigma_k^2}},
\end{equation}
where $F_k$, $V_k$ and $\sigma_k$ are the same as \autoref{eq:sigmae}; $R_k$ is the distance of $k$-th spaxel to the galaxy centre. We calculate these two quantities following \citet{Graham2018} as their code has been made publicly available\footnote{\url{https://github.com/marktgraham/lambdaR_e_calc}}, and the values have been corrected for PSF effects.

\section{Quality assessment}
\label{sec:quality_assessment}
The quality and characteristics of the data vary widely for a survey like MaNGA. For this reason, for a proper use of our catalogue, it is essential to be able to select subsets according to the quality of the models. In this section we discuss our qualitative assessment.

\subsection{Visual quality classification}
\label{sec:quality}
The MaNGA sample contains various types of galaxies, some of which, including merging galaxies, galaxy pairs, irregular galaxies, and strong bars, cannot be described by the dynamical models which assume a steady state. In other cases, the data quality is low because of low $S/N$ or strong dust absorptions. We thus grade the fitting quality by visually inspecting $V_{\rm rms}$ and $V$ maps recovered with the MFL model.

\begin{figure*}
    \centering
    \includegraphics[width=2\columnwidth]{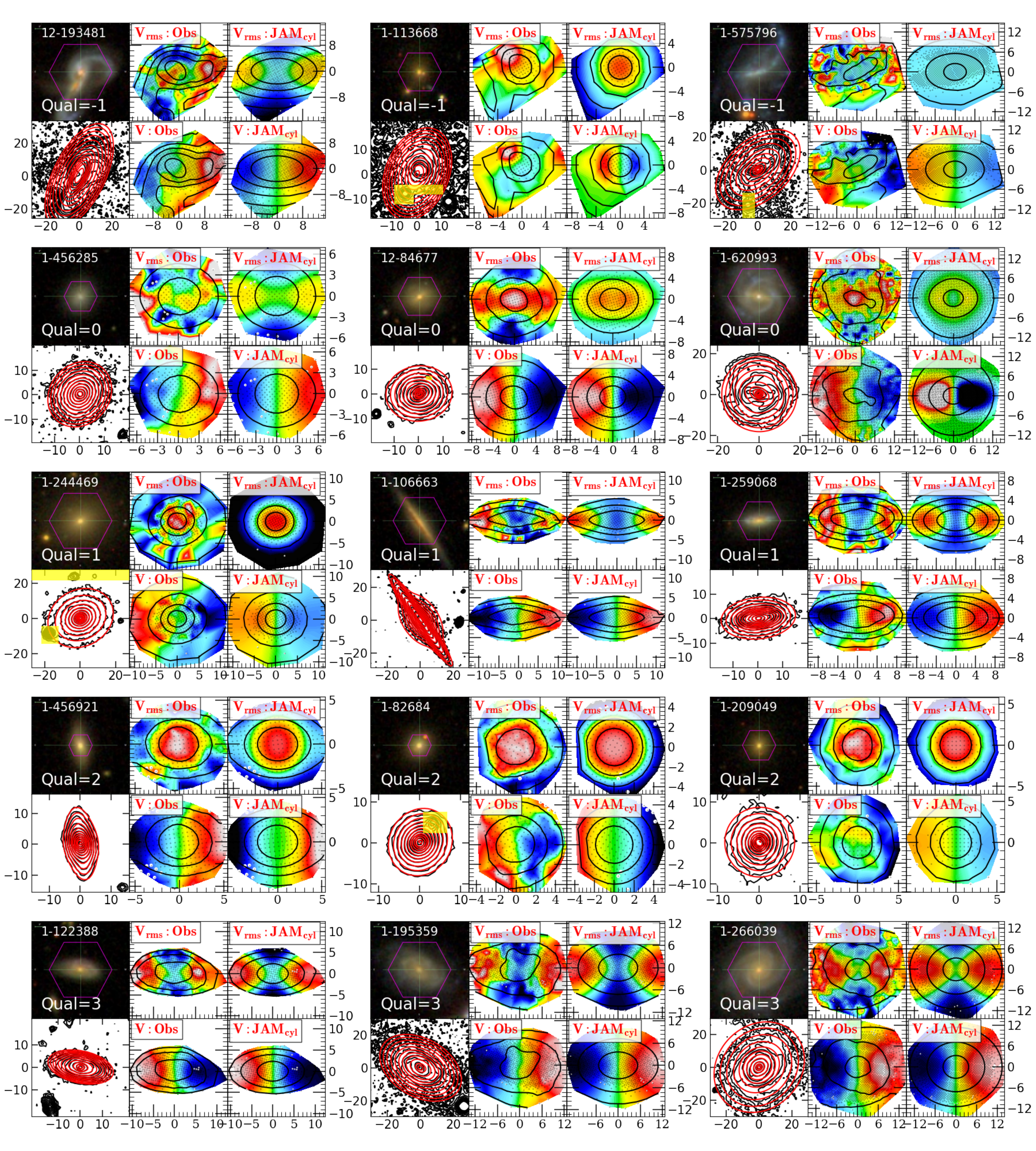}
    \caption{Examples of galaxies with different modelling qualities (from top to bottom: $\rm Qual=-1$, 0, 1, 2, and 3). For each galaxy, the RGB image, observed and modelled stellar kinematics (both $V_{\rm rms}$ and $V$), and SDSS $r$-band isophotes (black) overlaid with MGE contours (red) are presented. The contours, lines, and symbols are the same as \autoref{fig:kinpa_mgepa} and \autoref{fig:mge_example}.}
    \label{fig:quality}
\end{figure*}

We grade the modelling quality as $-1$, 0, 1, 2, 3 according to the following principles. $\rm Qual=-1$ means the galaxies' stellar kinematics are highly disturbed (e.g. merging galaxies, irregular galaxies or close galaxy pairs). One should not trust any of the kinematic/dynamic properties of these galaxies. $\rm Qual=0$ represents the galaxies with somewhat regular kinematic distributions but cannot be well-modelled by JAM, indicating inferior data quality (low S/N or low stellar velocity dispersion) or a problematic model (e.g. due to the presence of strong bars). $\rm Qual=1$ indicates an acceptable fit to the $V_{\rm rms}$ with the $V_{\rm rms}$ shape being somewhat predicted but the value being biased. $\rm Qual=2$ corresponds to the cases that have a good fit (both shape and value) to the $V_{\rm rms}$ but a bad fit to the line-of-sight velocity $V$ (see \autoref{sec:vlos} for the prediction of line-of-sight velocities). The highest quality $\rm Qual=3$ means that both $V_{\rm rms}$ and $V$ are well recovered. The numbers of galaxies in each quality group and the suggestions on the parameters than can be trusted in each quality group are given in \autoref{table:quality}. The example kinematic maps of different modelling qualities are shown in \autoref{fig:quality}. 

\begin{figure*}
    \centering
    \includegraphics[width=0.9\textwidth]{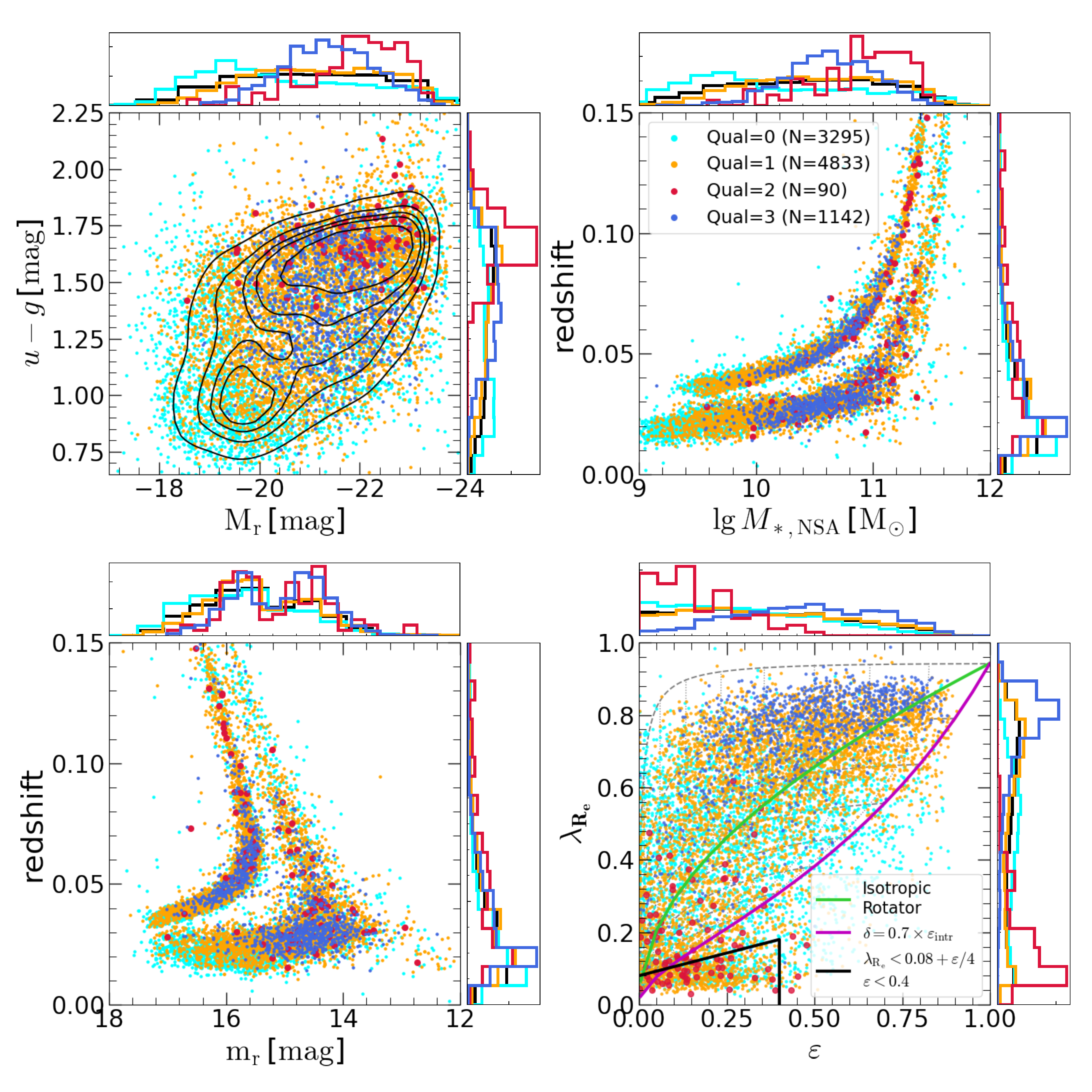}
    \caption{The distributions of MaNGA complete sample in the $u-g$ colour-magnitude (SDSS $r$-band) diagram (top left panel), the redshift-stellar mass diagram (top right panel), the redshift-apparent magnitude (SDSS r-band) diagram (bottom left panel), and the ($\lambda_{\rm R_e}$,$\varepsilon$) diagram (bottom right panel). Galaxies with different modelling qualities are shown with different colours (see the legend). The black contours are the distribution of the total sample. Histograms of each panel show the probability density functions for the whole sample (black) and each subsample of different modelling qualities. All the parameters are extracted from the NSA catalogue except for the ($\lambda_{\rm R_e}$,$\varepsilon$) diagram, where $\lambda_{\rm R_e}$ is calculated as \autoref{sec:sigma_lambda} and $\varepsilon$ is the observed ellipticity derived from the MGE models using the \textsc{mge\_half\_light\_isophote} software. In the ($\lambda_{\rm R_e}$,$\epsilon$) diagram, the green line represents the predicted relation for an edge-on ($i=90^{\circ}$) isotropic rotator from \citet{Binney2005} \citep[eq.~14]{Cappellari2016ARA&A}, while the magenta line denotes the edge-on relation from \citet{Cappellari2007} \citep[eq.~11]{Cappellari2016ARA&A}. The thin dotted lines show how the magenta line changes with different inclinations ($\Delta i=10^{\circ}$), while the thick dashed lines show how the galaxies move across the diagram with changing inclination for a set of given $\varepsilon_{\rm intr}$ values ($\Delta \varepsilon_{\rm intr}=0.1$). The lower-left region enclosed by the black solid lines ($\lambda_{\rm R_{\rm e}}<0.08+\varepsilon/4,\varepsilon<0.4$) define the region occupied by slow rotators \citep[eq.~19]{Cappellari2016ARA&A}.}
    \label{fig:colour-mag}
\end{figure*}

\begin{figure*}
    \centering
    \includegraphics[width=\textwidth]{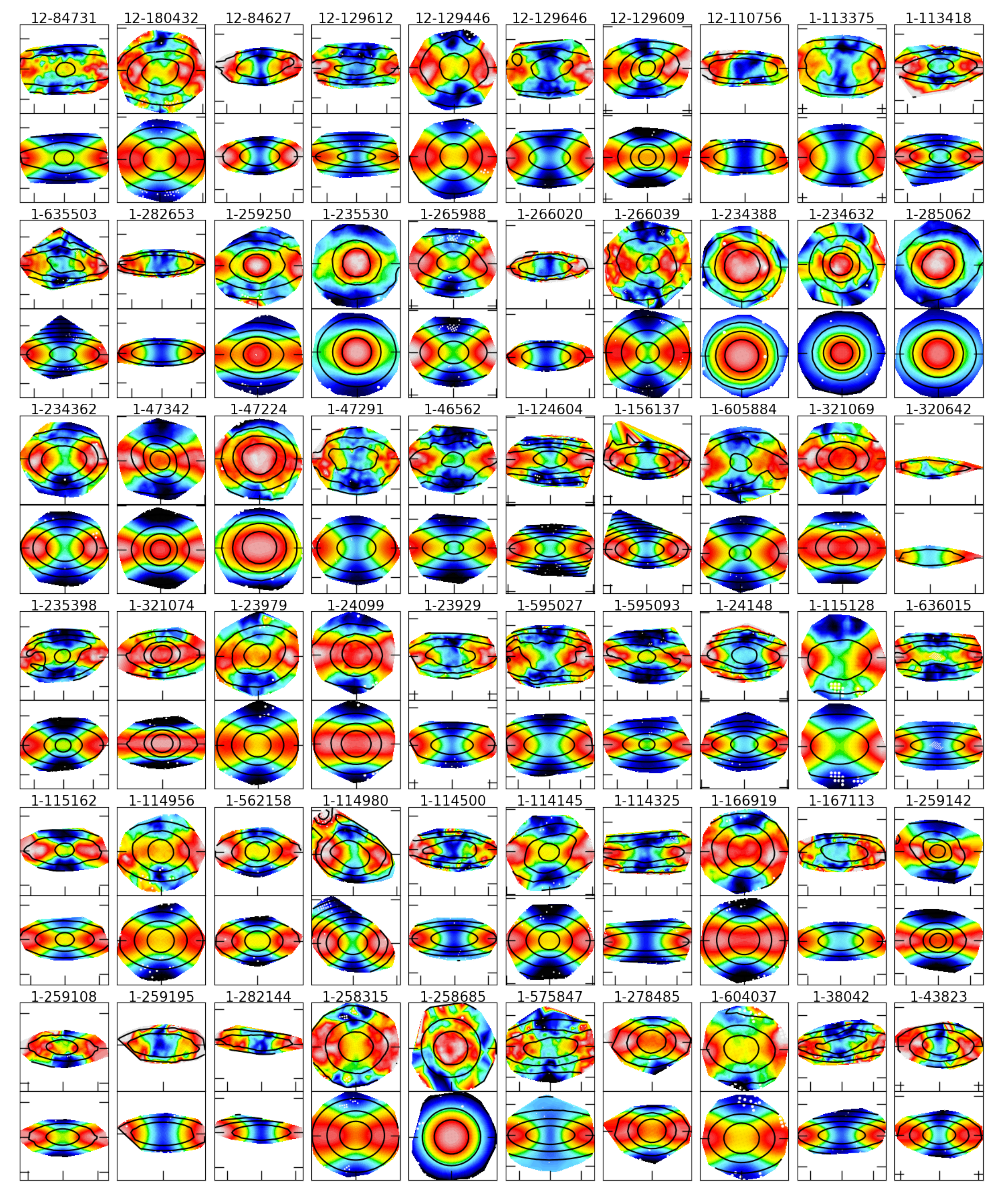}
    \caption{Examples of the mass-follows-light models for $\rm Qual=3$ galaxies, using the JAM$_{\rm cyl}$ method. For each galaxy, the top panel shows the observed $V_{\rm rms}$ map, overlaid with the black contours of observed surface brightness (in steps of 1 mag). In the bottom panel, the modelled $V_{\rm rms}$ map and the adopted MGE surface brightness (black contours) are shown. The white circles are the clipped bins, which are not included in the JAM fitting. Ticks are separated by $10^{\prime\prime}$. The JAM models (the observed and modelled $V_{\rm rms}$ maps) for the full sample can be found in \url{https://manga-dynpop.github.io}.}
    \label{fig:JAM_example}
\end{figure*}

In \autoref{fig:colour-mag}, we present the distributions of samples with different data qualities (except for $\rm Qual=-1$) on the colour-magnitude diagram, the redshift-stellar mass diagram, and the redshift-apparent magnitude diagram,  which are derived from the NASA-Sloan Atlas\footnote{\url{http://nsatlas.org/}} (NSA) catalogue \citep[see][]{Blanton2007,Blanton2011}. Moreover, we also present the ($\lambda_{\rm R_e}$,$\varepsilon$) diagram in the bottom right panel of \autoref{fig:colour-mag}. Compared to the parent sample, diverse distributions of stellar mass (or equivalently r-band absolute magnitude) are observed for different modelling qualities: the $\rm Qual=0$ sample contains a higher fraction of low mass galaxies, the distributions of $\rm Qual=1$ galaxies are flat, which is similar to the whole sample, while most of the $\rm Qual=2$ sample are massive galaxies. The stellar mass of $\rm Qual=3$ galaxies span in a range of $10^{10}-10^{11.3} M_{\odot}$. Note that the fitting quality has a strong dependence on the data quality of the spectrum. Some of the most massive ellipticals have low modelling quality because they have a low S/N spectrum (confirmed by the bottom left panel of \autoref{fig:colour-mag}: the modelling quality strongly depends on the apparent magnitude). The colour distributions are nearly identical to the whole sample except for the $\rm Qual=2$ sample, which has a higher fraction of red galaxies. In particular, most of $\rm Qual=2$ galaxies are massive and red, suggesting that they could be slow rotator galaxies and we confirm this in the ($\lambda_{\rm R_e}$, $\varepsilon$) diagram. As expected, the $\rm Qual=3$ galaxies contain more fast rotators compared to the parent sample, while the distributions of $\lambda_{\rm R_e}$ and $\varepsilon$ for $\rm Qual=0$ and $\rm Qual=1$ galaxies are very similar to the parent sample. We present some examples of the observed and modelled $V_{\rm rms}$ maps for the $\rm Qual=3$ galaxies in \autoref{fig:JAM_example}.

\begin{table*}
  \caption{Parameter guidance for MaNGA galaxies in different quality groups (see \autoref{sec:quality} for definition of the groups). From left to right, the columns are: (1) quality classification; (2) the number of galaxies; (3) the guidance for the use of quantities; (4) the reliable JAM-inferred parameters. For each quality (i.e. $\rm Qual=0, 1, 2, 3$), the reliable parameters for all galaxies in the given quality group (top) and for those galaxies satisfying the recommended selection criteria (bottom) are presented in the fourth column. Even for the reliable parameters for all galaxies (e.g. $M_{\rm T}(<R_{\rm e})$, $M_{\rm T}(<r_{1/2})$, $(M/L)_{\rm e}$, $\gamma_{_{\rm T}}$, $\overline{\gamma_{_{\rm T}}}$ for $\rm Qual\geqslant1$ galaxies), we recommend to further select the galaxies satisfying $|\lg(X_{\rm cyl}/X_{\rm sph})|<3\Delta$ (enclosed total masses or $M/L$) or $|X_{\rm cyl}-X_{\rm sph}|<3\Delta$ ( total density slopes) to remove the outliers. Here $X_{\rm cyl}$ and $X_{\rm sph}$ represent the quantities derived from $\rm JAM_{cyl}$ and $\rm JAM_{sph}$ models respectively. The $\Delta$ is the rms scatter of each quantity in different quality groups, which is taken from \autoref{tab:scatter}. The users can use more (or less) restrictive selection criteria to obtain purer (or more complete) samples based on their scientific purpose.} \setlength{\tabcolsep}{2mm}
\begin{tabular}{p{0.05\textwidth}p{0.05\textwidth}p{0.5\textwidth}p{0.35\textwidth}}
\hline
\hline
 Qual & $N_{\rm gal}$ & Guidance & Reliable parameters\\
\hline
$-1$ & 936 & No kinematic/dynamical properties should be used. & None\\
\hline
\multirow{4}{*}{0} & \multirow{4}{*}{3295} & \multirow{4}{0.5\textwidth}{$\sigma_{\rm e}$ and $\lambda_{R_{\rm e}}$ can be trusted. JAM-inferred parameters should be used with cautions: only the consistent $\rm JAM_{cyl}$ and $\rm JAM_{sph}$ inferred integrated quantities, i.e. the total mass within a sphere ($R_{\rm e}$ or $r_{1/2}$) or the total $M/L$ within $R_{\rm e}$, can be used (see \autoref{sec:uncertainty_mass}).} & $\sigma_{\rm e}$, $\lambda_{\rm R_{e}}$ \\
\cline{4-4}
& & & $M_{\rm T}(<R_{\rm e})$,$M_{\rm T}(<r_{1/2})$: $|\lg (X_{\rm cyl}/X_{\rm sph})|<3\Delta$\\
& & & $(M/L)_{\rm e}$: $|\lg (X_{\rm cyl}/X_{\rm sph})|<3\Delta$\\
& & & \\
\hline
\multirow{5}{*}{1} & \multirow{5}{*}{4833} & \multirow{5}{0.5\textwidth}{$\sigma_{\rm e}$ and $\lambda_{R_{\rm e}}$ can be trusted. Parameters related to the total mass distribution (e.g. total mass within $R_{\rm e}$, total mass-to-light ratios, and total density slopes) can be trusted. Parameters related to the decomposition between stellar and dark matter components (e.g. dark matter fraction within $R_{\rm e}$) should be used with cautions: only consistent $\rm JAM_{cyl}$ and $\rm JAM_{sph}$ inferred values can be trusted.} & $\sigma_{\rm e}$, $\lambda_{\rm R_{e}}$, $M_{\rm T}(<R_{\rm e})$, $M_{\rm T}(<r_{1/2})$, $(M/L)_{\rm e}$, $\gamma_{_{\rm T}}$, $\overline{\gamma_{_{\rm T}}}$ \\
\cline{4-4}
& & & $f_{\rm DM}(<R_{\rm e})$ : $|X_{\rm cyl}-X_{\rm sph}|<0.1$\\
& & & $M_{\ast}/L$: $|\lg (X_{\rm cyl}/X_{\rm sph})|<0.1$\\
& & & \\
& & & \\
\hline
\multirow{5}{*}{2} & \multirow{5}{*}{90} & \multirow{5}{0.5\textwidth}{$\sigma_{\rm e}$ and $\lambda_{R_{\rm e}}$ can be trusted. Parameters related to the total mass distribution (e.g. total mass within $R_{\rm e}$, total mass-to-light ratios and total density slopes) can be trusted. Parameters related to the decomposition between stellar and dark matter components (e.g. dark matter fraction within $R_{\rm e}$) should be used with cautions: only consistent $\rm JAM_{cyl}$ and $\rm JAM_{sph}$ inferred values can be trusted.} & $\sigma_{\rm e}$, $\lambda_{\rm R_{e}}$, $M_{\rm T}(<R_{\rm e})$, $M_{\rm T}(<r_{1/2})$, $(M/L)_{\rm e}$, $\gamma_{_{\rm T}}$, $\overline{\gamma_{_{\rm T}}}$ \\
\cline{4-4}
& & & $f_{\rm DM}(<R_{\rm e})$ : $|X_{\rm cyl}-X_{\rm sph}|<0.1$\\
& & & $M_{\ast}/L$: $|\lg (X_{\rm cyl}/X_{\rm sph})|<0.1$\\
& & & \\
& & & \\
\hline
3 & 1142 & All quantities are regarded as reliable. & All\\
\hline
\end{tabular}
\vspace{2mm}
\label{table:quality}
\end{table*}

\subsection{Predictions on the first velocity moments}

\label{sec:vlos}
The gravitational potential in a galaxy is only constrained by the even moment of the velocity. This is because, for a given density and tracer distributions and one can always revert the sense of rotation (change velocity sign) of arbitrary numbers of tracer stars, without having to change neither the density nor the tracer. For this reason, in our fitting process, as customary with JAM models, we only use the second velocity moments $V_{\rm rms}$ to determine the free parameters that define the galaxy shape and density distribution. To predict the projected second velocity moments, we do not need to make any assumption on how the second moment in the tangential direction $\overline{v_{\phi}^2}$, which appears in the Jeans \autoref{eq:JAMcyl} and \autoref{eq:JAMsph}, separates into ordered rotation and a random motion, defined by:
\begin{equation}
\label{eq:v_phi}
    \overline{v_{\phi}^2} = \overline{v_{\phi}}^2 + \sigma_{\phi}^2.
\end{equation}
To further obtain the line-of-sight velocity (i.e. the first velocity moment) map, one needs to make additional assumptions on the shape of velocity ellipsoid to evaluate $\overline{v_{\phi}}$. In this work, to quantify galaxy rotations we predict the velocities assuming an oblate velocity ellipsoid which satisfies
\begin{equation}
    \sigma_{\phi}^2 = \sigma_{\rm R}^2,\qquad \sigma_{\phi}^2 = \sigma_r^2,
\end{equation}
in the cylindrically or spherically aligned cases respectively, where $\sigma_{R}^2$ ($\sigma_{r}^2$) is determined from $\rm JAM_{cyl}$ ($\rm JAM_{sph}$) model, following \citet{Cappellari2008,Cappellari2020}. In the case of $\rm JAM_{sph}$, there are two natural choices for splitting ordered and random motions. Our choice corresponds to eq.~56 of \citet{Cappellari2020}, which is likely better suited for the fast rotator galaxies dominating our sample. After substituting $\sigma_{\phi}^2$ with $\sigma_{\rm R}^2$ ($\sigma_{r}^2$) in \autoref{eq:v_phi}, the intrinsic first velocity moment $\overline{v_{\phi}}$ is derived and can be used to predict modelled light-of-sight velocities $\overline{v_{\rm los}}$ (see section 3.1.5 of \citealt{Cappellari2008} for more details).

\begin{figure}
    \centering
    \includegraphics[width=1\columnwidth]{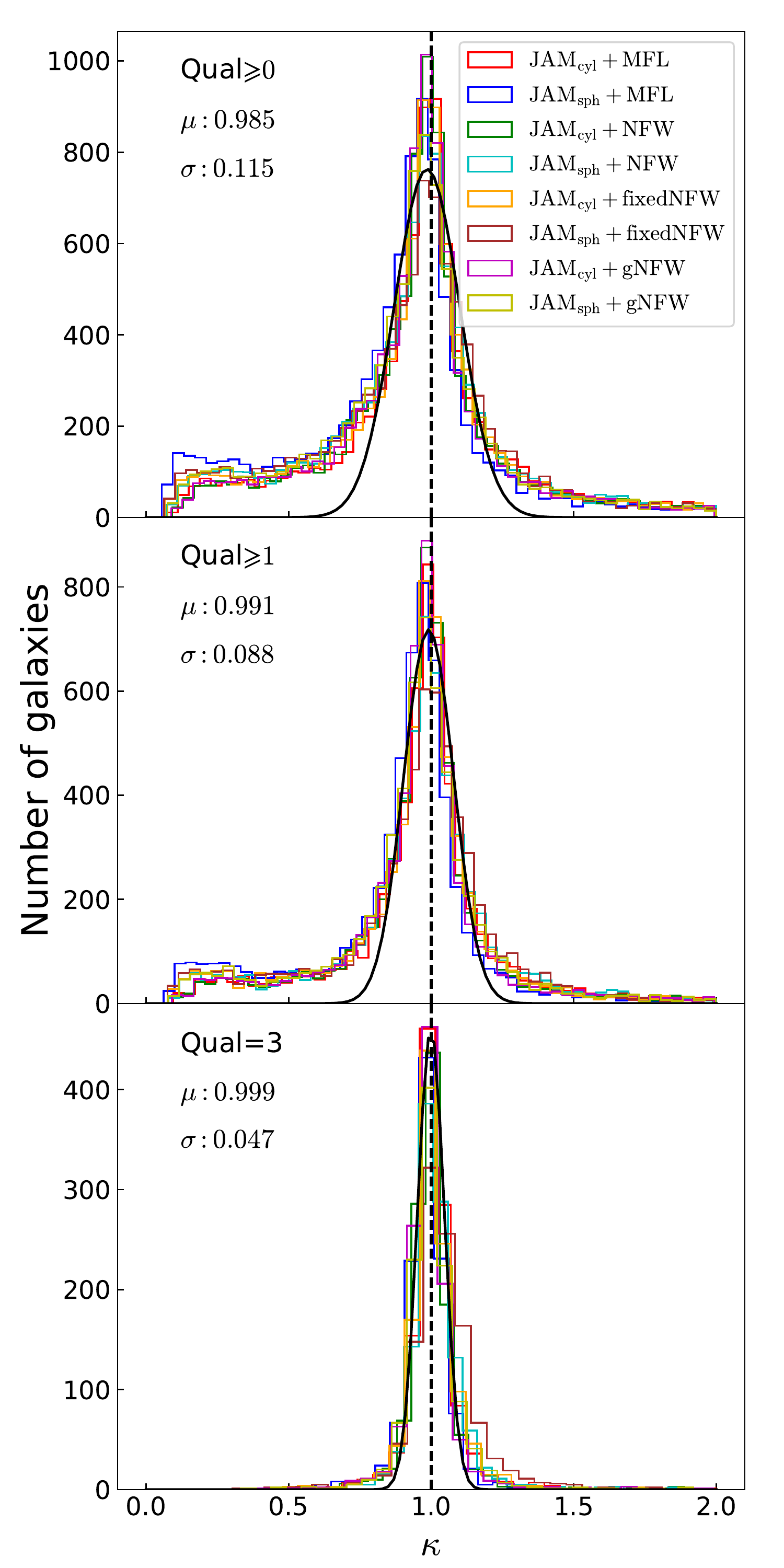}
    \caption{Distributions of the ratio between the observed and the predicted stellar velocities ($\kappa$; see \autoref{sec:vlos} for definition) for galaxies in different modelling quality groups (from top to bottom: $\rm Qual\geqslant 0$, $\rm Qual\geqslant 1$, and $\rm Qual=3$). In each panel, results for different JAM models are shown with histograms of different colours. The black solid curve is the Gaussian fit of the results for the JAM$_{\rm cyl}$ + MFL model (for the comparison with \citealt{Cappellari2016ARA&A}), with $\mu$ and $\sigma$ being the mean and standard deviation values of the Gaussian distribution.}
    \label{fig:kappa}
\end{figure}

To assess the validity of recovered line-of-sight velocity maps, we calculate the distribution of $\kappa$, defined as \citep[eq.~52]{Cappellari2008}
\begin{equation}
    \kappa = \frac{\sum_{k}F_{k}|x'_{k}V_{k}|}{\sum_{k}F_{k}|x'_{k}(\overline{v_{\rm los}})_{k}|},
\end{equation}
where $F_{k}$ is the flux of $k$-th Voronoi bin; $V_{k}$ and $(\overline{v_{\rm los}})_{k}$ are the corresponding observed and modelled line-of-sight velocities, respectively; $x'_{k}$ is the distance of the $k$-th Voronoi bin to the galaxy centre parallel to the major axis. In \autoref{fig:kappa}, we present the distributions of $\kappa$ for galaxies with different data qualities ($\rm Qual\geqslant0$, $\rm Qual\geqslant1$ and $\rm Qual=3$ from top to bottom). As can be seen, for all models, the distributions of $\kappa$ peak at $\kappa \approx 1$ regardless of the modelling quality, confirming that, after fitting the density, the first velocity moment can be well predicted under the assumption of an oblate velocity ellipsoid, as first noted in \citet{Cappellari2008} and quantified for a statistical sample of ETGs in \citet[fig.~11]{Cappellari2016ARA&A}. Here we find that this empirical fact is valid for all morphological galaxy types and for both JAM$_{\rm cyl}$ and JAM$_{\rm sph}$. The rms scatter of the $\kappa$ distribution decreases from low-quality samples to high-quality samples. If we exclude the $\rm Qual=0$ galaxies, the rms scatter becomes 8.8$\%$, compatible with the result of $\rm ATLAS^{3D}$ (see fig. 11 in \citealt{Cappellari2016ARA&A}). Specifically, the $\rm Qual=3$ galaxies, which are visually classified as the models that have good fit to the $V_{\rm rms}$ and $v_{\rm los}$, have a remarkably small observed rms scatter ($\sim 5\%$). 

\section{Realistic model uncertainties}
\label{sec:uncertainty}

A key strength of this project is that we derive mass models under two extreme  assumptions on the orientation of velocity ellipsoid, namely $\rm JAM_{cyl}$ and $\rm JAM_{sph}$. Moreover, for each galaxy in the sample and for each $\rm JAM_{cyl}$ and $\rm JAM_{sph}$ model we make 4 different assumptions for the dark/luminous mass decomposition, resulting in 8 different models to analyse the full sample of MaNGA galaxies. In this section, we use the differences between these different model assumptions to assess the level of systematics in our derived dynamical quantities. These dominate the formal (i.e. statistical) uncertainties, which are generally small, unreliable and not very useful for practical usage.

\subsection{$\rm JAM_{cyl}$ vs. $\rm JAM_{sph}$}

We focus on five quantities: the mass-weighted total density slope $\overline{\gamma_{_{\rm T}}}$, the dark matter fraction $f_{\rm DM}(<R_{\rm e})$, the dynamical mass-to-light ratio $(M/L)_{\rm e}$ and the total mass $M_{\rm T}(<R_{\rm e})$, all four quantities within a sphere of radius $R_{\rm e}$, and the reduced chi-square $\rm \chi^2/DOF$ (see \autoref{sec:cat} and \aref{sec:appendix_catalogue} for explanations of these quantities). To reduce the effect of bad fittings, we only use the galaxies of the highest quality ($\rm Qual=3$) in this test. We perform a linear fit to the quantities obtained with $\rm JAM_{cyl}$ and $\rm JAM_{sph}$, using the robust \textsc{lts\_linefit} \footnote{Version 5.0.19, from \url{https://pypi.org/project/ltsfit/}} procedure \citep{Cappellari2013a}, which combines the Least Trimmed Squares robust technique of \citet{Rousseeuw2006} into a least-squares fitting algorithm which allows for errors in all variables and intrinsic scatter.

\begin{figure*}
    \centering
    \includegraphics[width=1.95\columnwidth]{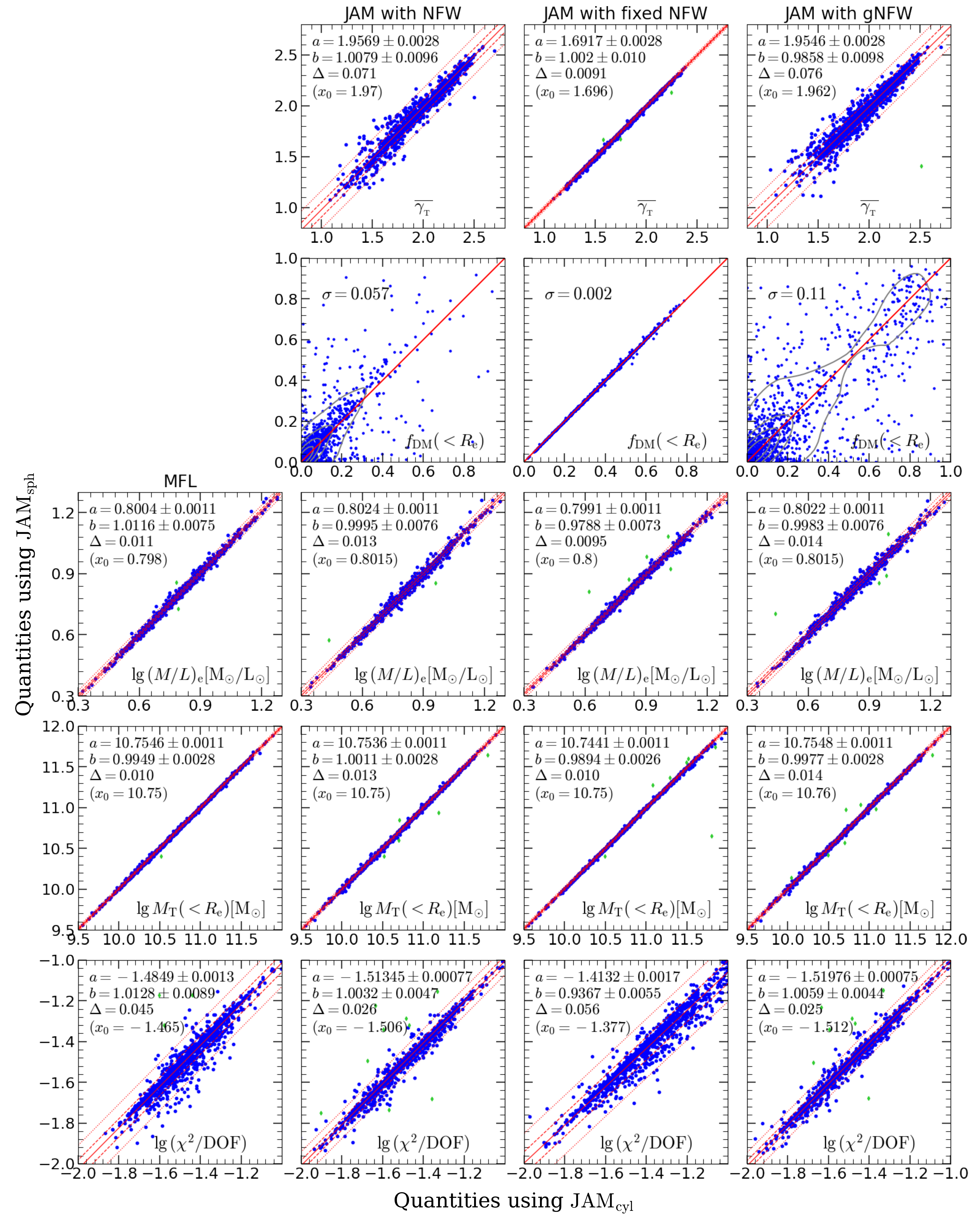}
    \caption{Comparisons of the results derived from JAM$_{\rm cyl}$ (X-axis) and from JAM$_{\rm sph}$ (Y-axis). The fitting is carried out with least-squares fitting on 1142 galaxies with $\rm Qual=3$. Fitting results for MFL, NFW, fixed NFW, and gNFW models (see \autoref{sec:model_design}) are shown from left to right. Four quantities, including the total mass slope $\overline{\gamma_{_{\rm T}}}$, dark matter fraction within $R_{\rm e}$, $f_{\rm DM}(<R_{\rm e})$, total mass within $R_{\rm e}$, $\lg\,M_{\rm T}(<R_{\rm e})$, and $\chi^2/\rm DOF$ are shown from top to bottom. Since the MFL model does not have stellar-DM decomposition, resulting in the equivalent total mass density slope and luminosity density slope, we do not show $\overline{\gamma_{_{\rm T}}}$ and $f_{\rm DM}(<R_{\rm e})$ for the MFL model. In each panel except for the panels of $f_{\rm DM}(<R_{\rm e})$, a linear fit is performed to the parameters derived with different velocity ellipsoid assumptions (i.e. JAM$_{\rm cyl}$ and JAM$_{\rm sph}$), using the \textsc{lts\_linefit} software \citep{Cappellari2013a} with \texttt{clip=6}. The fitting results are listed in the upper left of each panel. The solid, dashed, and dotted red lines represent the best-fit, 1$\sigma$ (68$\%$ confidence level) scatter and 2.6$\sigma$ (99$\%$ confidence level) scatter, respectively. The green symbols are the detected outliers beyond 6$\sigma$ confidence level. In each panel of the second rows, the grey contour is a kernel density estimate of the galaxy distribution (using \href{https://docs.scipy.org/doc/scipy/reference/generated/scipy.stats.gaussian_kde.html}{scipy.stats.gaussian\_kde}), while the red line is the one-to-one relation. We do not fit the data of $f_{\rm DM}(<R_{\rm e})$ panels. In those cases, the $\sigma$ represents the 68th percentile of the absolute difference $|f_{\rm DM, cyl}-f_{\rm DM, sph}|$.}
    \label{fig:JAMcyl_JAMsph_qual3}
\end{figure*}

The results are presented in \autoref{fig:JAMcyl_JAMsph_qual3}. In the bottom panels, the reduced chi-square $\rm \chi^2/DOF$ of $\rm JAM_{cyl}$ and $\rm JAM_{sph}$ are statistically indistinguishable (with observed rms scatter $\Delta = 0.025$ dex) for NFW and gNFW models, slightly larger than the observed rms scatter ($\Delta = 0.022$ dex) using $\rm ATLAS^{3D}$ data \citep{Cappellari2020}. The observed rms scatter slightly increases to $\Delta = 0.045$ dex for the MFL model and $\Delta = 0.056$ dex for the fixed NFW model, consistent with the expectation that the difference in $\rm \chi^2/DOF$ between $\rm JAM_{cyl}$ and $\rm JAM_{sph}$ is smaller for the more flexible mass models.

As shown in the third and fourth rows of \autoref{fig:JAMcyl_JAMsph_qual3}, the enclosed dynamical mass-to-light ratio $(M/L)_{\rm e}$ and the total mass within effective radius $M_{\rm T}(<R_{\rm e})$ are reliable quantities that are nearly unaffected by the assumption of the orientation of velocity ellipsoids. For the MFL model, the rms scatter of $(M/L)_{\rm e}$ is $\Delta = 0.011$ dex, indicating an error of $\Delta/\sqrt{2} = 1.8\%$ in the individual $(M/L)_{\rm e}$. The errors of NFW and gNFW models are $\sim 2.1\%$ ($\Delta \approx 0.013 \rm\, dex$), and the smallest errors are found in the models with fixed NFW dark halo ($\sim 1.6\%$, $\Delta \approx 0.0095 \rm\, dex$).

The mass-weighted total density slopes $\overline{\gamma_{_{\rm T}}}$ are also very consistent in $\rm JAM_{cyl}$ and $\rm JAM_{sph}$. The observed rms scatters of NFW, fixed NFW, and gNFW models are $\Delta=0.071$, $\Delta=0.0091$, and $\Delta=0.076$, respectively. The values of observed scatter for NFW and gNFW models are close to those obtained from $\rm ATLAS^{3D}$ ($\Delta=0.094$, \citealt[fig.~12]{Cappellari2020}), reconfirming the validity of the approach. 

When the dark halo slope approaches the slope of the stellar density, a change in the stellar $M/L$ becomes indistinguishable from an increase in the dark matter fraction. For this reason, one cannot expect to be able to uniquely constrain the $f_{\rm DM}(<R_{\rm e})$ in every individual galaxy. Nonetheless, $\rm JAM_{cyl}$ and $\rm JAM_{sph}$ inferred $f_{\rm DM}(<R_{\rm e})$ values are statistically consistent, with 1$\sigma$ scatter of 0.057, 0.0020, and 0.11 for NFW, fixed NFW, and gNFW models, where the 1$\sigma$ scatter is defined as 68th percentile of the absolute difference $|f_{\rm DM, cyl}-f_{\rm DM, sph}|$. However, for some extreme cases, the difference can be up to 0.7 between the two different JAM assumptions, which implies the $f_{\rm DM}(<R_{\rm e})$ is essentially unconstrained by the data without more restrictive priors. One can also find that different mass models predict different $f_{\rm DM}(<R_{\rm e})$, which will be discussed in more detail in \autoref{sec:fdm_NFW_gNFW}. Therefore, when future users analyse dark matter fractions with our catalogue, it is important to select galaxies with consistent $f_{\rm DM}(<R_{\rm e})$ values for different models. 

\subsection{Uncertainties of density slope, enclosed $M/L$ and  mass}
\label{sec:uncertainty_mass}

\begin{figure*}
    \centering
    \includegraphics[width=2\columnwidth]{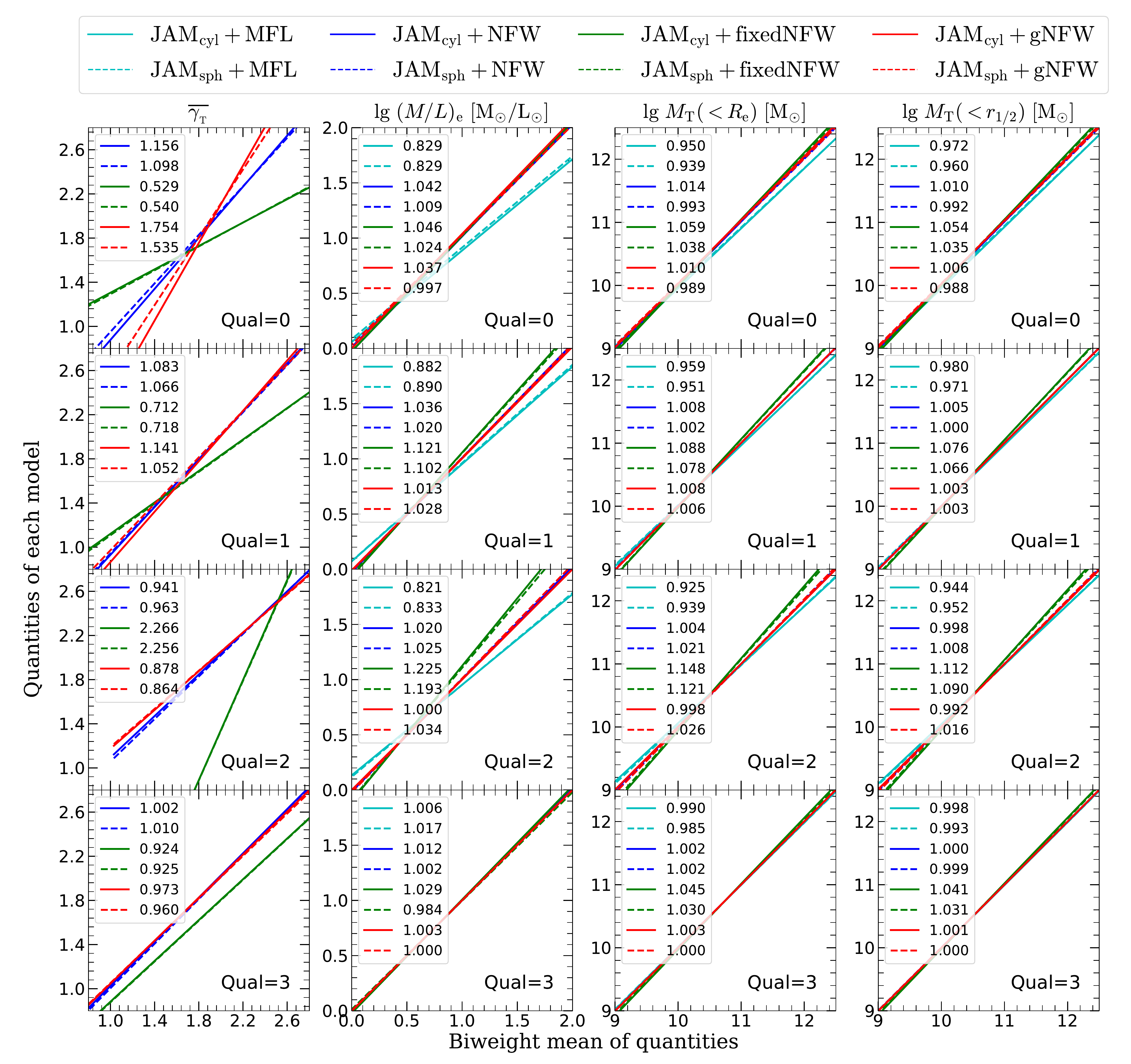}
    \caption{The comparisons of quantities between each model and the biweight mean value of eight models (except for the total density slope $\overline{\gamma_{_{\rm T}}}$ which does not account for the MFL model) for different quality groups, with best-fit straight-lines of each model which are derived from \textsc{lts\_linefit} software. Results of galaxies with quality 0, 1, 2, and 3 are shown from top to bottom, and four quantities, including the total density slope $\overline{\gamma_{_{\rm T}}}$, the dynamical mass-to-light ratio $(M/L)_{\rm e}$, the total mass within 2D effective radius $M_{\rm T}(<R_{\rm e})$, and the total mass within 3-dimensional half-light radius $M_{\rm T}(<r_{1/2})$ are shown from left to right. The colours correspond to different mass models, while solid and dashed lines represent the JAM${_{\rm cyl}}$ and JAM${_{\rm sph}}$ methods (see the top legends for the details). Corresponding slopes of best-fit straight lines for each model are shown in the legend of each panel.}
    \label{fig:denslope_MLdyn_MtRe_Mtrhalf_slopes}
\end{figure*}

\begin{figure*}
    \centering
    \includegraphics[width=2\columnwidth]{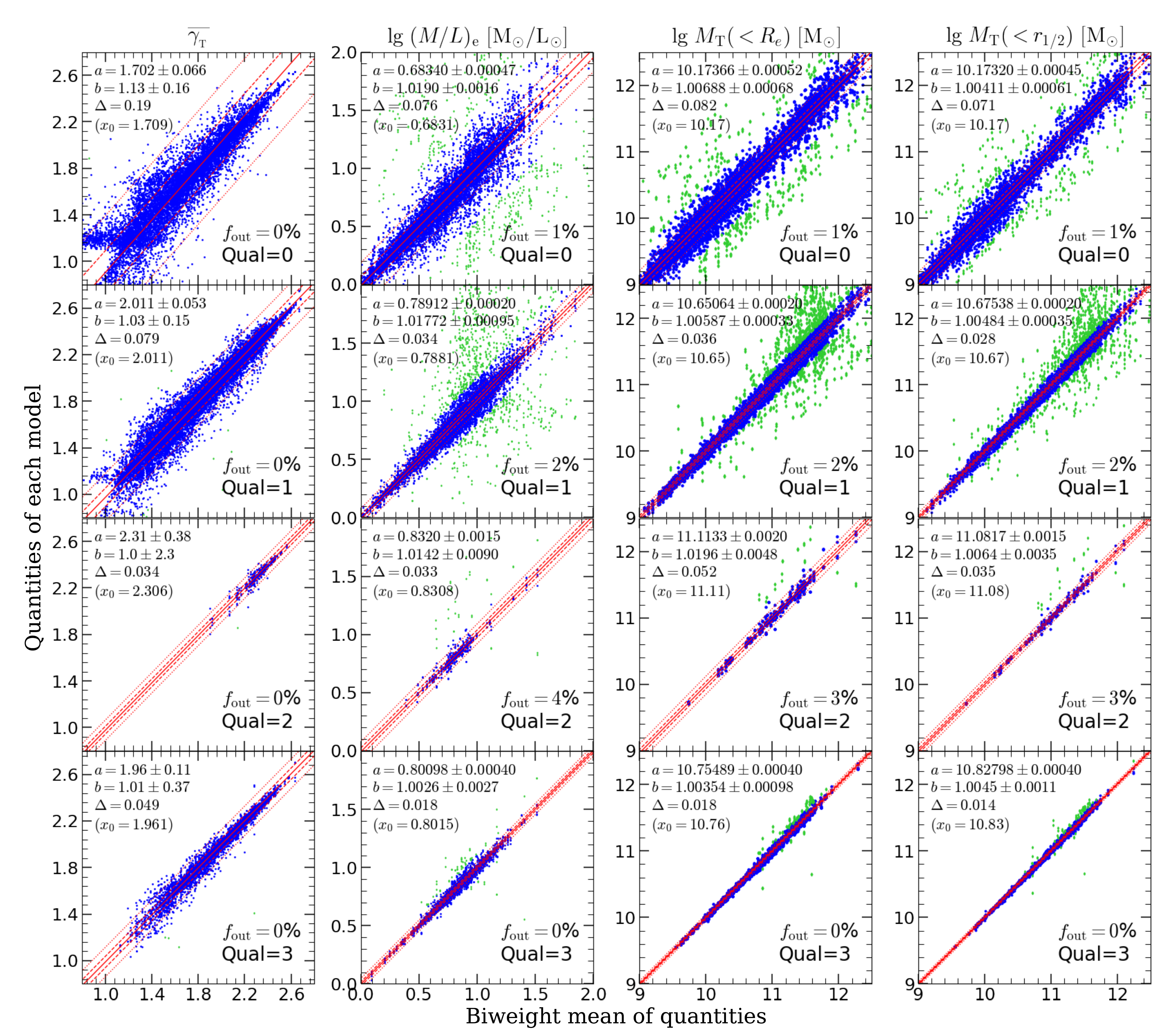}
    \caption{Systematic measurement uncertainties for galaxies in different quality groups. The panels are the same as \autoref{fig:denslope_MLdyn_MtRe_Mtrhalf_slopes}. In each panel, the Y-axis represents the corresponding quality of all 8 models ($\overline{\gamma_{_{\rm T}}}$ is not shown for MFL and fixed NFW models, since it is uniquely decided by luminosity density slope in the MFL model and values of the fixed NFW model deviate far from other models as shown in \autoref{fig:denslope_MLdyn_MtRe_Mtrhalf_slopes}), and the X-axis is for the biweight mean value of the quantities (see \autoref{sec:uncertainty_mass} for definition) for different models. Using \textsc{lts\_linefit} procedure with \texttt{clip=6}, the best-fit, 1$\sigma$ (68$\%$ of values) scatter and 2.6$\sigma$ (99$\%$ of values) scatter are fitted and shown with red solid, dashed and dotted lines. The green symbols are the detected outliers beyond 6$\sigma$ confidence level, while the fraction of the detected outliers are listed in each panel. }
    \label{fig:denslope_MLdyn_MtRe_Mtrhalf_errors}
\end{figure*}

We present the scatter of parameters among different models in \autoref{fig:denslope_MLdyn_MtRe_Mtrhalf_slopes}. The scatter is calculated with respect to the biweight mean \citep[pg.~417]{Hoaglin1983} of different models. We compare the values of each model with the biweight mean value, which are shown in \autoref{fig:denslope_MLdyn_MtRe_Mtrhalf_slopes} and \autoref{fig:denslope_MLdyn_MtRe_Mtrhalf_errors}. The biweight mean is known as a robust method to determine the central location of a distribution, which is shown to be more robust compared to the conventional mean value \citep{Beers1990,Andrews2015} in statistics with outliers or small statistical size. With the comparisons between models, we aim to investigate the systematic bias between different models and measure the systematic uncertainties of derived quantities.

On the left column of \autoref{fig:denslope_MLdyn_MtRe_Mtrhalf_slopes}, the comparisons of mass-weighted total density slopes $\overline{\gamma_{_{\rm T}}}$ between 6 models are shown (the mass-follows-light models are excluded because that their total density slopes are uniquely decided by their luminosity density slopes, which do not vary with JAM models). For the NFW and gNFW models, the slopes of best-fit lines are close to unity for $\rm Qual\geqslant1$ galaxies, while deviations from unity are observed for $\rm Qual=0$ galaxies. The values of the fixed NFW model deviate from biweight mean values regardless of quality, indicating that the underlying theoretical assumption for the halo is unable to accurately predict the real galaxies. Moreover, the uncertainties of $\overline{\gamma_{_{\rm T}}}$ between NFW and gNFW models are presented on the left column of \autoref{fig:denslope_MLdyn_MtRe_Mtrhalf_errors} (the fixed NFW model is excluded because they are a theoretical prediction, which we found is not sufficiently reproducing real galaxies). From $\rm Qual=0$ to $\rm Qual=3$, the observed rms scatter $\Delta$ ranges from 0.19 to 0.049. For the galaxies with $\rm Qual>0$, the observed scatters ($\Delta=0.079$ for $\rm Qual=1$, $\Delta=0.034$ for $\rm Qual=2$, $\Delta=0.049$ for $\rm Qual=3$) are compatible with the values $\Delta = 0.13$ obtained by \citet{Poci2017} and $\Delta=0.094$ obtained by \citet{Cappellari2020}. For our alternative definition of total density slope or \autoref{eq:density2}, i.e. the logarithmic total density slope $\gamma_{_{\rm T}}$, the rms scatters are nearly identical ($\Delta=0.18$ for $\rm Qual=0$, $\Delta=0.082$ for $\rm Qual=1$, $\Delta=0.047$ for $\rm Qual=2$, $\Delta=0.049$ for $\rm Qual=3$; \autoref{tab:scatter}).

The comparisons of dynamical mass-to-light ratio $(M/L)_{\rm e}$, total mass within effective radius $M_{\rm T}(<R_{\rm e})$, and total mass within a sphere of 3D half-light radius $M_{\rm T}(<r_{1/2})$ are presented in the second, third, and fourth columns of \autoref{fig:denslope_MLdyn_MtRe_Mtrhalf_slopes} and \autoref{fig:denslope_MLdyn_MtRe_Mtrhalf_errors}. In \autoref{fig:denslope_MLdyn_MtRe_Mtrhalf_slopes}, for the $\rm Qual=0$ galaxies, $(M/L)_{\rm e}$ measurements of the MFL model are systematically smaller compared to other models especially at the high $(M/L)_{\rm e}$ end. With increasing quality, the systematic difference between different mass models becomes indistinguishable (the correlations between models have a slope of nearly unity except for the $\rm Qual=2$ galaxies with a significantly small number). 

The excellent agreement between MFL and more flexible halo models, with good-quality data, is consistent with the same JAM results for ETGs in ATLAS$^{\rm 3D}$ \citep[fig.~9]{Cappellari2013a}. While the smaller $M/L$ for MFL models on low-quality data is consistent with the LEGA-C JAM modelling \citep[fig.~9]{vanHoudt2021}. This is important: Given that with good data the MFL models return the correct $(M/L)_{\rm e}$, the difference observed at low $S/N$ is unlikely to be a genuine effect. Instead, it must be due to the more flexible halo models being unable to correctly constrain the dark halo and returning overestimated $f_{\rm DM}(<R_{\rm e})$ and consequently too large $(M/L)_{\rm e}$. The tendency of low-quality data to cause overestimated $f_{\rm DM}(<R_{\rm e})$ was noted in \citet[fig.~10]{Cappellari2013a} and we observe also in this work. We verify that this is not due to galaxies with better data having smaller data coverage and smaller dark matter fraction. This suggests that, in general, the MFL models provide a more robust estimate of the $(M/L)_{\rm e}$ than more general models, even in the presence of $M/L$ gradients, and should be preferred with inferior data quality.

Following popular practice, we also compare enclosed masses from different models. However, we stress that these comparisons do not contain any information that is not already better visible in the $(M/L)_{\rm e}$ plots. In fact, by definition $M_{\rm T}(<R_{\rm e})=L(<R_{\rm e})\times(M/L)_{\rm e}$. This implies that the mass plots can be obtained by multiplying both $x-y$ axes by the same luminosity, effectively stretching the axes' scale and making any differences more difficult to detect. This explains the fact that masses have the same scatter as the $M/L$ within the numerical uncertainties of the clipping process. This is perhaps the reason why masses are often a preferred way of comparing dynamical results than $M/L$.
No apparent systematic differences are observed for the $M_{\rm T}(<R_{\rm e})$, regardless of quality (third column in \autoref{fig:denslope_MLdyn_MtRe_Mtrhalf_slopes}). The scatters of $(M/L)_{\rm e}$ and $M_{\rm T}(<R_{\rm e})$ are nearly identical: $\Delta=0.082$ dex for $\rm Qual=0$, $\Delta=0.036$ dex for $\rm Qual=1$, $\Delta=0.052$ dex for $\rm Qual=2$, and $\Delta=0.018$ dex for $\rm Qual=3$ (\autoref{fig:denslope_MLdyn_MtRe_Mtrhalf_errors}). The larger scatters for $\rm Qual=2$ galaxies than those for $\rm Qual=1$ galaxies may be due to the significantly smaller number of $\rm Qual=2$ galaxies. Moreover, the scatter of $M_{\rm T}(<R_{\rm e})$ for $\rm Qual=3$ galaxies ($\Delta=0.018$ dex) is compatible with $\Delta=0.037$ dex obtained from the full sample of $\rm ATLAS^{3D}$ \citep{Cappellari2013a}, suggesting that models of $\rm Qual=3$ galaxies are at the same level of accuracy as those of $\rm ATLAS^{3D}$. The total mass within 3-dimensional half-light radius $M_{\rm T}(<r_{1/2})$ is a robust quantity that has no systematic bias between models (\autoref{fig:denslope_MLdyn_MtRe_Mtrhalf_slopes}), with a small observed scatter ranging from $\Delta=0.071$ dex to $\Delta=0.014$ dex in different quality groups (\autoref{fig:denslope_MLdyn_MtRe_Mtrhalf_errors}).

The observed scatters $\Delta$ and corresponding errors are presented in \autoref{tab:scatter}. In summary, the quantities related to the total mass distribution are reliable and no systematic bias between models is observed for the $\rm Qual\geqslant1$ galaxies (except for the $\overline{\gamma_{_{\rm T}}}$ inferred from the fixed NFW model). For the $\rm Qual=0$ galaxies, the total mass within a sphere of $R_{\rm e}$ or $r_{1/2}$ can still be trusted. The observed rms scatters of quantities significantly decrease with increasing modelling quality, again confirming the importance of our visual quality control and the usefulness of using multiple models to estimate realistic parameter uncertainties. Furthermore, the small scatter also suggests the insensitivity of these quantities to different mass models (i.e. MFL, NFW, fixed NFW, and gNFW models) and different assumptions on the orientation of the velocity ellipsoid (i.e. JAM$_{\rm cyl}$ and JAM$_{\rm sph}$).

\begin{table*}
\centering
\caption{The errors of individual quantities ($\overline{\gamma_{_{\rm T}}}$, $\gamma_{_{\rm T}}$, $(M/L)_{\rm e}$, $M_{\rm T}(<R_{\rm e})$, $M_{\rm T}(<r_{1/2})$) for different quality groups. The slope and $\Delta$ denote the slope and observed scatter obtained by the \textsc{lts\_linefit} procedure. The error is defined as $\Delta/\sqrt{2}$ assuming that the quantities on both axes are comparable. The errors of quantities in this table are derived from eight models listed in \autoref{tab:model}, except for the errors of $\overline{\gamma_{_{\rm T}}}$ and $\gamma_{_{\rm T}}$, for which MFL and fixed NFW models are not accounted for (see the text in \autoref{sec:uncertainty_mass} for more explanations).}
\label{tab:scatter}
\setlength{\tabcolsep}{2.0mm}{
    \begin{tabular}{|c|c|c|c|c|c|c|c|c|c|c|c|c|c|}
        \hline
        \multicolumn{1}{c|}{\multirow{2}*{Quantities}} &\multicolumn{3}{|c}{Qual=0}&\multicolumn{3}{|c}{$\rm Qual=1$}&\multicolumn{3}{|c}{$\rm Qual=2$}&\multicolumn{3}{|c}{$\rm Qual=3$}&\multirow{2}*{Mass models} \\
        \cline{2-13}
        \multicolumn{1}{c|}{} & slope & $\Delta$ & error & slope & $\Delta$ & error & slope & $\Delta$ & error & slope & $\Delta$ & error\\
        \hline
        $\overline{\gamma_{_{\rm T}}}$ & 1.13 & 0.19 & 0.13 & 1.03 & 0.079 & 0.056 & 1.00 & 0.034 & 0.024 & 1.01 & 0.049 & 0.035 & NFW, gNFW\\
        $\gamma_{_{\rm T}}$ & 1.10 & 0.18 & 0.13 & 1.02 & 0.082 & 0.058 & 1.00 & 0.047 & 0.033 & 1.01 & 0.049 & 0.035 & NFW, gNFW\\
        $(M/L)_{\rm e}$ & 1.02 & 0.076 dex & 13.2\% & 1.02 & 0.034 dex & 5.69\% & 1.01 & 0.033 dex & 5.52\% & 1.00 & 0.018 dex & 2.97\% & All\\
        $M_{\rm T}(<R_{\rm e})$ & 1.01 & 0.082 dex & 14.3\% & 1.01 & 0.036 dex & 6.04\% & 1.02 & 0.052 dex & 8.84\% & 1.00 & 0.018 dex & 2.97\% & All\\
        $M_{\rm T}(<r_{1/2})$ & 1.00 & 0.071 dex & 12.3\% & 1.00 & 0.028 dex & 4.66\% & 1.01 & 0.035 dex & 5.86\% & 1.00 & 0.014 dex & 2.31\% & All\\
        \hline
    \end{tabular}}
\end{table*}

\subsection{Model uncertainties on dark matter fractions}
\label{sec:fdm_NFW_gNFW}

Dynamical modelling in general, can only measure total densities, or equivalently enclosed masses. As far as we currently understand, gravity does not distinguish between luminous, baryonic or dark matter. For this reason, it is obvious that if one were to allow part of the dark matter to be distributed like the baryons, any dark matter decomposition would become degenerate. This implies that the decomposition of the total density into baryonic and dark matter always necessarily involves some level of assumptions. In practice, we know from numerical simulations, e.g. EAGLE \citep{EAGLE} and IllustrisTNG \citep{TNG1,TNG2,TNG3,TNG4,TNG5}, that dark matter is expected to be more extended and smooth than the stars and this allows us to place some constraints on its contribution. However, these assumptions do not remove all degeneracies and make dark matter a much more uncertain and assumption-dependent parameter than the other dynamical quantities we can more directly measure.

Using the galaxies of best quality ($\rm Qual=3$), the consistency of $f_{\rm DM}(<R_{\rm e})$ between $\rm JAM_{cyl}$ and $\rm JAM_{sph}$ is already shown in the second rows of \autoref{fig:JAMcyl_JAMsph_qual3}. Although there is no systematic differences between the inferred values of $\rm JAM_{cyl}$ and $\rm JAM_{sph}$, the non-negligible fraction of outliers makes it necessary to exclude the cases with significant inconsistency when the readers want to use the DM-stellar decomposition results of our modelling.

\begin{figure*}
    \centering
    \includegraphics[width=\textwidth]{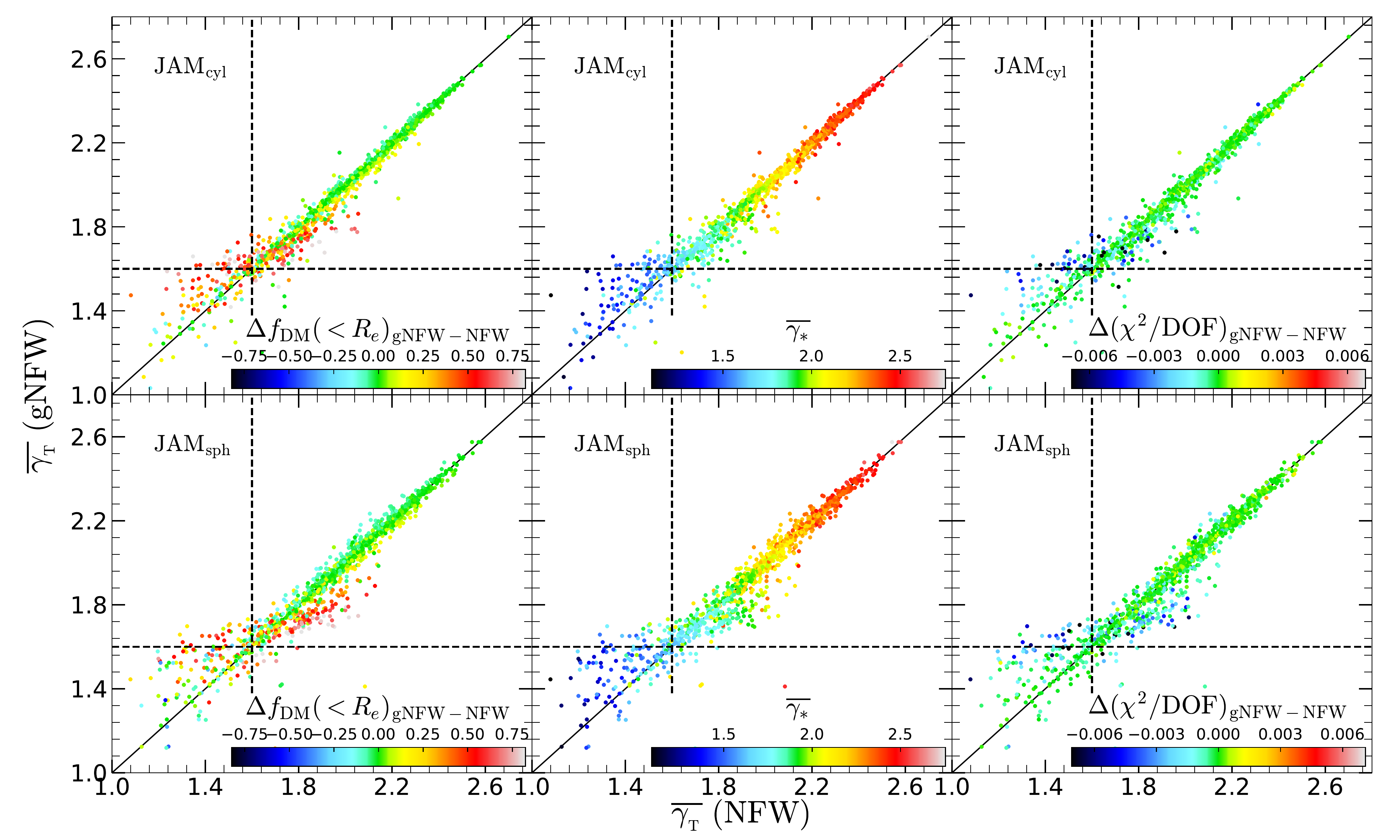}
    \caption{The comparisons of mass-weighted total density slope $\overline{\gamma_{_{\rm T}}}$ derived from NFW models (X-axis) and gNFW models (Y-axis) for galaxies with $\rm Qual=3$, colour-coded by dark matter difference of the two models ($\Delta f_{\rm DM}(<R_{\rm e})_{\rm gNFW-NFW}$, left panels), stellar density slope ($\overline{\gamma_{\ast}}$, middle panels), and $\chi^2$ difference between two models ($\Delta (\chi^2/\rm DOF)_{\rm gNFW-NFW}$, right panels). Results for JAM$_{\rm cyl}$ and JAM$_{\rm sph}$ are shown in the top and bottom panels, respectively. In each panel, the vertical and horizontal black dashed lines represent the mass-weighted density slope of 1.6, which is approximately the steepest dark matter density slope (i.e. with the largest absolute value) allowed by the gNFW density profiles (see \autoref{tab:model}, but note that the mass-weighted slope always has a positive sign by definition, as opposed to the inner slope parameter $\gamma$ of gNFW profiles).}
    \label{fig:fdmRe_NFW_gNFW_qual3}
\end{figure*}

However, we also notice that there are still some cases in \autoref{fig:JAMcyl_JAMsph_qual3} that have significant differences between different mass models even after excluding the outliers. To understand the origin of the systematic offset of $f_{\rm DM}(<R_{\rm e})$, we present \autoref{fig:fdmRe_NFW_gNFW_qual3}, in which $\rm Qual=3$ galaxies are plotted on the $\overline{\gamma_{_{\rm T}}} ({\rm NFW})-\overline{\gamma_{_{\rm T}}} ({\rm gNFW})$ plane, colour-coded by the dark matter fraction difference between NFW and gNFW models, the stellar density slope, and the $\chi^2/\rm DOF$ difference between NFW and gNFW models from left to right. Results for JAM$_{\rm cyl}$ and JAM$_{\rm sph}$ are shown in the top and bottom panels, respectively. As can be seen, most galaxies have nearly identical total density slopes and indistinguishable differences in dark matter fraction under different mass models (i.e. the NFW model vs. the gNFW model), while a subset of galaxies has slightly different $\overline{\gamma_{_{\rm T}}}$ and significantly inconsistent $f_{\rm DM}(<R_{\rm e})$. As seen from the right panels of \autoref{fig:fdmRe_NFW_gNFW_qual3}, the $\rm \chi^2/DOF$ values of the gNFW model are smaller than those of the NFW model for this subset of galaxies, which is expected due to more free parameters in the gNFW model. For this reason, we use $\overline{\gamma_{_{\rm T}}}$ of the gNFW model as the reference to investigate the origin of the differences in $\overline{\gamma_{_{\rm T}}}$ and $f_{\rm DM}(<R_{\rm e})$ between the two models.

For the galaxies with shallow total and stellar density slopes ($1<\overline{\gamma_{_{\rm T}}}<1.6$ and $1<\overline{\gamma_{_{\ast}}}<1.6$), the variable inner density slope of gNFW profile allows $\overline{\gamma_{_{\rm DM}}}$ to be similar to $\overline{\gamma_{_{\ast}}}$ or even steeper. Thus, the dark matter fraction of the gNFW model can be large for the outliers with 1 < $\overline{\gamma_{_{\ast}}}$ < $\overline{\gamma_{_{\rm T}}}$ < 1.6. But for the NFW model, due to the shallower $\overline{\gamma_{_{\rm DM}}}$, the only way to reach the same total density slope as the gNFW model is to reduce the dark matter fraction. However, it is still impossible to reach a steeper $\overline{\gamma_{_{\rm T}}}$ than $\overline{\gamma_{_{\ast}}}$ for the NFW model even with very small dark matter fraction, leading to the much smaller $f_{\rm DM}(<R_{\rm e})$ and slightly shallower $\overline{\gamma_{_{\rm T}}}$ of the NFW model (see the left panels of \autoref{fig:fdmRe_NFW_gNFW_qual3}).

Meanwhile, the stellar density slope $\overline{\gamma_{_{\ast}}}$ is always steeper than $\overline{\gamma_{_{\rm DM}}}$ for the galaxies with $1.6<\overline{\gamma_{_{\rm T}}}<2$. Since the total density slope $\overline{\gamma_{_{\rm T}}}$ is determined by the interplay between shallower dark matter density slope $\overline{\gamma_{_{\rm DM}}}$ and steeper stellar density slope $\overline{\gamma_{_{\ast}}}$, the dark matter fraction in the gNFW model can be higher responding to the steeper $\overline{\gamma_{_{\rm DM}}}$. For the galaxies with steeper total density slopes ($\overline{\gamma_{_{\rm T}}}>2$), no distinguishable differences in $\overline{\gamma_{_{\rm T}}}$ and $f_{\rm DM}(<R_{\rm e})$ between NFW and gNFW models are observed.

\section{Summary}
\label{sec:sum}
In this work, we construct a full catalogue of dynamical quantities for the complete sample of 10K galaxies with integral-field kinematics from the MaNGA survey (SDSS DR17; \citealt{Abdurro'uf2022}) with their fitting quality carefully assessed. The quantities are derived using a detailed dynamical model (JAM; \citealt{Cappellari2008,Cappellari2020}), which is based on the axisymmetric Jeans equations and had been demonstrated by others to be more accurate than other more-general techniques in recovering the total mass distribution. Four mass models (i.e. MFL, NFW, fixed NFW, and gNFW; see \autoref{sec:model_design} for details) and two assumptions on the orientation of velocity ellipsoid (i.e. the cylindrically-aligned $\rm JAM_{cyl}$ and the spherically-aligned $\rm JAM_{sph}$ velocity ellipsoids; see \autoref{sec:JAM}) are adopted in this work. By fitting the observed second velocity moments $V_{\rm rms}$, the free parameters of different models are optimised to find the maximum likelihood value. Based on the comparison between observed and modelled $V_{\rm rms}$ and $V$ maps, the sample is visually classified into different modelling quality groups ($\rm Qual=-1$, 0, 1, 2, 3; see \autoref{sec:quality}).

The main results of tests on the robustness of measured quantities (e.g. mass-weighted total density slope $\overline{\gamma_{_{\rm T}}}$, dynamical mass-to-light ratio $(M/L)_{\rm e}$, enclosed total mass within a sphere of $R_{\rm e}$ $M_{\rm T}(<R_{\rm e})$, enclosed total mass within a sphere of $r_{1/2}$ $M_{\rm T}(<r_{1/2})$; see \autoref{sec:cat} for details) and the systematic uncertainties in different models are: 
\begin{itemize}[align=left,leftmargin=2em,itemsep=1em]
    \item As shown in \autoref{fig:kappa}, the first order velocity moments can be well recovered ($\kappa \approx 1$) under the assumption of oblate velocity ellipsoid, regardless of modelling quality. The rms scatter of $\kappa$ distribution decreases with the improvement of modelling quality and has a remarkably small scatter of 0.05 for the $\rm Qual=3$ galaxies.
    \item The comparisons between quantities inferred from $\rm JAM_{cyl}$ and $\rm JAM_{sph}$ in \autoref{fig:JAMcyl_JAMsph_qual3} show that no systematic offsets between these two methods. The small observed scatter of $\overline{\gamma_{_{\rm T}}}$ ($\Delta \approx 0.071-0.076$), $f_{\rm DM}(<R_{\rm e})$ (1$\sigma$ scatter of 0.057 for the NFW model, 1$\sigma$ scatter of 0.11 for the gNFW model), $M_{\rm T}(<R_{\rm e})$ ($\Delta \approx 0.010-0.014$ dex), and $\rm \chi^2/DOF$ ($\Delta \approx 0.025-0.045$ dex) are comparable to the values obtained in \citet{Cappellari2020}, reconfirming the validity of these two approaches.
    \item Systematic bias and errors are explored by comparing measured quantities of eight different models (four mass models with different assumptions on the dark matter halos and two assumptions on the orientation of velocity ellipsoid for each mass model). No distinguishable systematic differences in $\overline{\gamma_{_{\rm T}}}$ (MFL and fixed NFW models are excluded), ${(M/L)_{\rm e}}$, $M_{\rm T}(<R_{\rm e})$ and $M_{\rm T}(<r_{1/2})$ are observed for different modelling qualities (the slopes of best-fit straight-lines in \autoref{fig:denslope_MLdyn_MtRe_Mtrhalf_slopes}  are close to 1). The enclosed masses computed with MFL models are highly consistent with those computed with more flexible models with DM, when the data are good. However, models with DM can provide significantly larger enclosed masses on low-quality data. The fixed NFW model inferred $\overline{\gamma_{_{\rm T}}}$ values systematically deviate from other models due to its fixed-halo assumption, thus are not recommended for use. The systematic errors for the galaxies of different modelling qualities are listed in \autoref{tab:scatter}. The small observed rms scatters shown in \autoref{fig:denslope_MLdyn_MtRe_Mtrhalf_errors} suggest that the quantities related to total mass distribution are reliable for $\rm Qual\geqslant1$ galaxies. Specifically, the enclosed total mass within a sphere of $R_{\rm e}$ or $r_{1/2}$ for $\rm Qual=0$ galaxies can be used by selecting the consistent values between models.
    \item The dark matter fraction $f_{\rm DM}(<R_{\rm e})$ are consistent between different models for most galaxies. However, a systematic offset of $f_{\rm DM}(<R_{\rm e})$ between different mass models is reported for a subset of galaxies (left columns in \autoref{fig:fdmRe_NFW_gNFW_qual3}). The inconsistency for this subset of galaxies is due to the fact that the dark matter profile of gNFW is more flexible in contributing to the total density profile as discussed in \autoref{sec:fdm_NFW_gNFW}.
\end{itemize}

Thanks to the large sample of MaNGA survey, this catalogue provides robust dynamical modelling for $\sim 10000$ galaxies, which makes it the largest catalogue of galaxies with dynamical properties so far. The MaNGA survey consists of different types of nearby galaxies, providing an unbiased and representative sample to study the relations related to stellar dynamics (e.g. fundamental plane, mass-size plane, total density slopes, dark matter fractions, IMF variations). Furthermore, we expect that this catalogue solely or combined with stellar population analysis, e.g. FIREFLY \citep{Goddard2017,Neumann2022}, Pipe3D \citep{Sanchez2022}, and pPXF \citep[][Paper II]{Lu2023}, will bring new insights into our understanding of galaxy formation and evolution. In the following papers of this project, we will present a catalogue of stellar population properties \citep[][Paper II]{Lu2023}, the dynamical scaling relations \citep[][Paper III]{Zhu2023}, the combined analysis with weak gravitational lensing \citep[][Paper IV]{Wang2023}, and the IMF variations (Lu et al. in preparation).

\section*{Acknowledgements}
We acknowledge Dr. Dandan Xu for helpful discussions on this paper. This work is partly supported by the National Key Research and Development Program of China (No. 2018YFA0404501 to SM), by the National Science Foundation of China (Grant No. 11821303, 11761131004 and 11761141012). This project is also partly supported by Tsinghua University Initiative Scientific Research Program ID 2019Z07L02017. We also acknowledge the science research grants from the China Manned Space Project with NO. CMS-CSST-2021-A11. KZ and RL acknowledge the support of National Nature Science Foundation of China (Nos 11988101,11773032,12022306), the support from the Ministry of Science and Technology of China (Nos. 2020SKA0110100),  the science research grants from the China Manned Space Project (Nos. CMS-CSST-2021-B01,CMS-CSST-2021-A01), CAS Project for Young Scientists in Basic Research (No. YSBR-062), and the support from K.C.Wong Education Foundation.

Funding for the Sloan Digital Sky 
Survey IV has been provided by the 
Alfred P. Sloan Foundation, the U.S. 
Department of Energy Office of 
Science, and the Participating 
Institutions. 

SDSS-IV acknowledges support and 
resources from the Center for High 
Performance Computing  at the 
University of Utah. The SDSS 
website is www.sdss.org.

SDSS-IV is managed by the 
Astrophysical Research Consortium 
for the Participating Institutions 
of the SDSS Collaboration including 
the Brazilian Participation Group, 
the Carnegie Institution for Science, 
Carnegie Mellon University, Center for 
Astrophysics | Harvard \& 
Smithsonian, the Chilean Participation 
Group, the French Participation Group, 
Instituto de Astrof\'isica de 
Canarias, The Johns Hopkins 
University, Kavli Institute for the 
Physics and Mathematics of the 
Universe (IPMU) / University of 
Tokyo, the Korean Participation Group, 
Lawrence Berkeley National Laboratory, 
Leibniz Institut f\"ur Astrophysik 
Potsdam (AIP),  Max-Planck-Institut 
f\"ur Astronomie (MPIA Heidelberg), 
Max-Planck-Institut f\"ur 
Astrophysik (MPA Garching), 
Max-Planck-Institut f\"ur 
Extraterrestrische Physik (MPE), 
National Astronomical Observatories of 
China, New Mexico State University, 
New York University, University of 
Notre Dame, Observat\'ario 
Nacional / MCTI, The Ohio State 
University, Pennsylvania State 
University, Shanghai 
Astronomical Observatory, United 
Kingdom Participation Group, 
Universidad Nacional Aut\'onoma 
de M\'exico, University of Arizona, 
University of Colorado Boulder, 
University of Oxford, University of 
Portsmouth, University of Utah, 
University of Virginia, University 
of Washington, University of 
Wisconsin, Vanderbilt University, 
and Yale University.
\section*{Data Availability}
The analysis results (including the catalogue and the figures of model fitting) are publicly available as supplementary files on the journal website. The catalogue is a single FITS file ($\sim$20 MB), while the data model is presented in \aref{sec:appendix_catalogue}. The full data release including the catalogue, more supplementary files (e.g. the mass profiles), and the updates (if any) on the catalogue will be posted in \url{https://manga-dynpop.github.io}. The MaNGA kinematics data is publicly available in \url{https://www.sdss4.org/dr17/manga/manga-data/data-access/} and the corresponding imaging data is available in \url{https://www.sdss.org/dr12/imaging/images/}.

\section*{Software Citations}
This work uses the following software packages:

\begin{itemize}

\item
\href{https://github.com/astropy/astropy}{{Astropy}}
\citep{astropy1, astropy2}

\item
\href{https://pypi.org/project/pymultinest/}{{Pymultinest}}
\citep{Buchner2014}

\item
\href{https://github.com/matplotlib/matplotlib}{{Matplotlib}}
\citep{Matplotlib2007}

\item
\href{https://github.com/numpy/numpy}{{NumPy}}
\citep{Numpy2011,Numpy2020}

\item
\href{https://www.python.org/}{{Python}}
\citep{Python3}

\item
\href{https://github.com/scikit-image/scikit-image}{{Scikit-image}}
\citep{scikit-image}

\item
\href{https://github.com/scipy/scipy}{{Scipy}}
\citep{Scipy2020}

\item
\href{https://pypi.org/project/jampy/}{{JamPy}}
\citep{Cappellari2008,Cappellari2020}

\item
\href{https://pypi.org/project/mgefit/}{{MgeFit}}
\citep{Cappellari2002}

\item
\href{https://pypi.org/project/pafit/}{{PaFit}}
\citep{Krajnovic2006}

\item
\href{ https://pypi.org/project/ltsfit/}{{LtsFit}}
\citep{Cappellari2013a}

\end{itemize}

\bibliographystyle{mnras}
\bibliography{ref}


\appendix
\section{Measuring the axial ratio of dust ring}
\label{sec:appendix_dustring}
\begin{figure*}
    \centering
    \includegraphics[width=\textwidth]{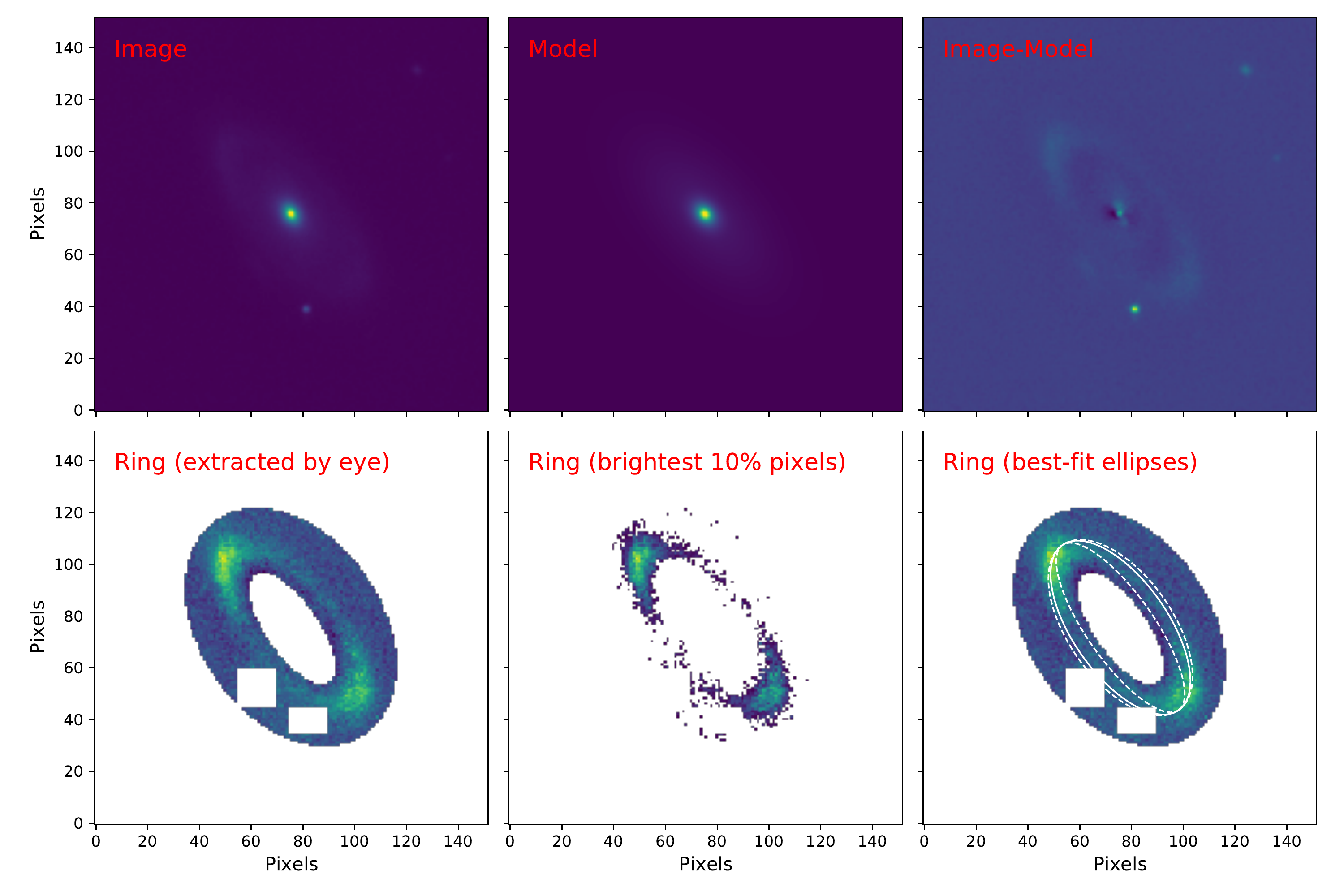}
    \caption{An example of measuring the axial ratio of the dust ring for the galaxy 8341-6101. From left to right and top to bottom, the image, the MGE model, the residual map (image-model), the ring structure extracted by eye, the brightest $10\%$ pixels of the extracted ring structure which are used to obtain the best-fit ellipse, and the ring structure overlaid with best-fit ellipses. The white solid line in the bottom right panel is the best-fit ellipse, while the white dashed lines are the ellipses which have $10\%$ larger residuals than the best-fit one.}
    \label{fig:extract_ring}
\end{figure*}

We perform MGE fitting on the images of the dust ring galaxies and derive the residual maps by subtracting the MGE models from the images. The ring structure can not be modelled by MGE (the surface brightness of Gaussians always decreases with increasing radius, while the ring structure results in a bump on the surface brightness profile), thus the ring structure is presented in the residual map. Then we extract the ring by: 1) roughly selecting an ellipse shell that contains the ring structure by eye, 2) selecting the brightest $5\%-20\%$ (typically 10$\%$) pixels within the ellipse shell, and 3) masking the foreground stars by hand if necessary. We use the least-squares estimator for 2D ellipses, \href{https://scikit-image.org/docs/stable/api/skimage.measure.html}{skimage.measure.EllipseModel} \citep[scikit-image,][]{scikit-image}, to obtain the best-fit ellipse model and then iteratively increase/decrease the axial ratio of the ellipse, with the center, the semi-minor axis, and the position angle of the best-fit ellipse fixed, until the residuals increase by 10$\%$ (see \autoref{fig:extract_ring}). In this way, we are able to obtain the upper and the lower limit of the observed axial ratio of the dust ring, with which the range of galaxy
inclination can be calculated. 

\section{Properties included in the catalogue}
\label{sec:appendix_catalogue}
We present the data model of this catalogue in \autoref{tab:cat}, which lists the dynamical properties derived from the analysis described in this paper.
\onecolumn
\begin{longtable}{cllll}
\caption{Dynamical properties for each model. For each property, the Header Data Unit (HDU) in which it is stored, the name, its units, and a brief description are presented. The HDU ranging from 2 to 9 correspond to models $\rm JAM_{cyl} + MFL$, $\rm JAM_{sph} + MFL$, $\rm JAM_{cyl} + NFW$, $\rm JAM_{sph} + NFW$, $\rm JAM_{cyl} + fixed NFW$, $\rm JAM_{sph} + fixed NFW$, $\rm JAM_{cyl} + gNFW$, $\rm JAM_{sph} + gNFW$. The NFW profile is written as $\rho_{_{\rm DM}}(r) = \rho_s\left(\frac{r}{r_s}\right)^{-1}\left(\frac{1}{2}+\frac{1}{2}\frac{r}{r_s}\right)^{-2}$,
while the gNFW profile is written as
$\rho_{_{\rm DM}}(r) = \rho_s\left(\frac{r}{r_s}\right)^{\gamma}\left(\frac{1}{2}+\frac{1}{2}\frac{r}{r_s}\right)^{-\gamma-3}$. The properties with prefix `nsa\_' are taken from the NSA catalogue \citep{Blanton2007,Blanton2011}.}
\label{tab:cat}\\
\hline \hline
\shortstack{HDU \\ (1)} & \shortstack{Name \\ (2)} & \shortstack{Units \\ (3)} & \shortstack{Description \\ (4)}\\
\hline
\endfirsthead

\multicolumn{5}{c}{\autoref{tab:cat} -- continued}\\
\hline \hline
\shortstack{HDU \\ (1)} & \shortstack{Parameters \\ (2)} &  \shortstack{Units \\ (3)} & \shortstack{Description \\ (4)}\\
\hline
\endhead

\hline \hline \multicolumn{5}{c}{{Continued on next page}}
\endfoot

\hline \hline
\endlastfoot
0 & Primary &  & Empty primary header\\
\hline
1 & plate&  & The plate ID (e.g. 7443)\\
  & ifudsgn &  & The IFU design ID (e.g.12703)\\
  & plateifu &  & The plate+ifudsgn name (e.g. 7443-12703)\\
  & mangaid &  & Unique MaNGA ID (e.g. 1-114145)\\
  & obj\_ra &  degree & Right ascension of the science object in J2000\\
  & obj\_dec &  degree & Declination of the science object in J2000\\
  & ebvgal &  & E(B-V) value from sdss dust routine for this IFU\\
  & target &  & Flag for subsample of MaNGA (Primary: 0, Secondary: 1, colour-Enhanced: 2)\\
  & rmax\_arcsec & arcsec & The kinematic data range, which is defined as the largest radius of the Voronoi bins\\
  & DA &  Mpc & Adopted angular-diameter distance,\\&&&\ \ \ \ \ \    with a flat Universe of $\Omega_{\rm m}=0.307$, $h=0.677$ \citep{Planck2016}\\
  & Re\_arcsec\_MGE & arcsec & Effective radius (projected circular half-light radius from MGE fitting, \\&&&\ \ \ \ \ \ in SDSS r-band)\\
  & Rmaj\_arcsec\_MGE  & arcsec & Major axis of elliptical half-light isophote from MGE fitting, in SDSS r-band\\
  & Lum\_tot\_MGE  & $\rm \lg(L_{\odot})$& Total luminosity from MGE fitting, in SDSS r-band, not corrected for \\&&&\ \ \ \ \ \  the Galactic and internal dust extinction\\
  & Lambda\_Re &  & Specific stellar angular momentum within elliptical half-light isophote, beam corrected\footnote{Following eq. 5 of \citet{Graham2018}}\\
  & Sigma\_Re  & $\rm km\,s^{-1}$ & Effective velocity dispersion within elliptical half-light isophote\\
  & Eps\_MGE &  & Ellipticity of the half-light isophote from MGE fitting\\
  & PA\_phot & degree & The photometric position angle (PA\footnote{The standard astronomical PA measured counter-clockwise from the image Y-axis (assumed to coincide with North).}) measured from MGE fitting, in SDSS r-band\\
  & PA\_kin & degree & The kinematic PA measured from MaNGA velocity field\\
  & PA\_kin\_flag & & The flag for kinematic PA (0 for unreliable, 1 for reliable)\\
  & nsa\_iauname &  & The accepted IAU name\\
  & z &  & Redshift of the galaxy\\
  & nsa\_field &  & The SDSS field covering the target\\
  & nsa\_run &  & The SDSS run covering the target\\
  & nsa\_camcol &  & The SDSS camcol covering catalogue position\\
  & nsa\_version &  & The version of the NSA catalogue used to select these targets\\
  & nsa\_id &  & The NSAID field in the NSA catalogue v1\\
  & nsa\_nsaid\_v1b &  & The NSAID of the target in the NSA\_v1b\_0\_0\_v2 catalogue (if applicable)\\
  & nsa\_sersic\_absmag &  & Absolute magnitude estimates for FNugriz from K-corrections ($\Omega_{\rm m}=0.3$,\\&&&\ \ \ \ \ \  $\Omega_{\rm \Lambda}=0.7$, $ h=1$), the value is interpreted as $M-5\lg h$\\
  & nsa\_elpetro\_absmag &  & As nsa\_sersic\_absmag but from elliptical Petrosian apertures \\
  & nsa\_sersic\_mass & $\rm \lg(h^{-2} \ M_{\odot})$ & Stellar mass from K-correction fit for Sersic fluxes\\
  & nsa\_elpetro\_mass  & $\rm \lg(h^{-2} \ M_{\odot})$ & Stellar mass from K-correction fit for elliptical Petrosian fluxes\\
  & nsa\_sersic\_ba &  & Axial ratio b/a from 2D Sersic fit in SDSS r-band\\
  & nsa\_sersic\_n &  & Sersic index from 2D Sersic fit in SDSS r-band\\
  & nsa\_sersic\_phi  & degree & Angle (E of N) of major axis in 2D Sersic fit (r-band)\\
  & nsa\_sersic\_th50  & arcsec & Sersic 50\% light radius along major axis (r-band) \\
  & nsa\_sersic\_flux & nanomaggies & 2D Sersic fit flux in FNugriz (GALEX-SDSS photometric systems)\\
  & Qual &  & Visual quality of JAM models, classified as -1, 0, 1, 2, 3 (from worst to best)\\
  & drp3qual &  & Data reduction quality marked by DRP pipeline, \\&&&\ \ \ \ \ \    1 for high-quality, 0 for critical-quality or unusual quality\\
 \hline
2 & inc\_deg &  degree & Best-fit inclination angle (being 90$^{\circ}$ for edge-on)\\
($\rm JAM_{cyl}+$
  & beta\_z &  & Best-fit radial velocity anisotropy in cylindrical coordinates\\
MFL)  & log\_ML\_dyn &  $\rm \lg(M_{\odot}/L_{\odot})$ & Best-fit dynamical mass-to-light ratio\\
  & kappa &  & The ratio between modelled line-of-sight velocity field and the observed one\\
  & log\_Mt\_Re  & $\rm \lg(M_{\odot})$ & Enclosed total mass within a sphere of effective radius\\
  & chi2\_dof &  & The reduced chi-square of the best-fit model (The values are scaled to \\&&&\ \ \ \ \ \  account for the effect of standard deviation of the $\rm \chi^2$ itself, \\&&&\ \ \ \ \ \ should be only used in the comparison between different models)\\
  & rhalf\_arcsec  & arcsec & Radius of the sphere which encloses half the total luminosity\\
  & log\_Mt\_rhalf  & $\rm \lg(M_{\odot})$ & Enclosed total mass within a sphere of 3D half-light radius\\
  & MW\_Gt\_Re &  & Mass-weighted total density slope within a sphere of effective radius\\
  & MW\_Gt\_rhalf &  & Mass-weighted total density slope within a sphere of 3D half-light radius\\
  & Gt\_Re &  & Average logarithmic total density slope between 0.1 and 1 effective radius\\
\hline
3 & inc\_deg  & degree & Best-fit inclination angle (being 90$^{\circ}$ for edge-on)\\
($\rm JAM_{sph}+$ 
  & beta\_r  & & Best-fit radial velocity anisotropy in spherical coordinates\\
MFL) & log\_ML\_dyn  & $\rm \lg(M_{\odot}/L_{\odot})$ & Best-fit dynamical mass-to-light ratio\\
  & kappa &  & The ratio between modelled line-of-sight velocity field and the observed one\\
  & log\_Mt\_Re  & $\rm \lg(M_{\odot})$ & Enclosed total mass within a sphere of effective radius\\
  & chi2\_dof &  & The reduced chi-square of the best-fit model (The values are scaled to \\&&&\ \ \ \ \ \ account for the effect of standard deviation of the $\rm \chi^2$ itself, \\&&&\ \ \ \ \ \ should be only used in the comparison between different models)\\
  & rhalf\_arcsec  & arcsec & Radius of the sphere which encloses half the total luminosity\\
  & log\_Mt\_rhalf  & $\rm \lg(M_{\odot})$ & Enclosed total mass within a sphere of 3D half-light radius\\
  & MW\_Gt\_Re &  & Mass-weighted total density slope within a sphere of effective radius\\
  & MW\_Gt\_rhalf &  & Mass-weighted total density slope within a sphere of 3D half-light radius\\
  & Gt\_Re &  & Average logarithmic total density slope between 0.1 and 1 effective radius\\
\hline
4 & inc\_deg &  & Best-fit inclination angle (being 90$^{\circ}$ for edge-on)\\
($\rm JAM_{cyl}+$ 
  & beta\_z &  & Best-fit radial velocity anisotropy in cylindrical coordinates\\
NFW) & log\_ML\_stellar  & $\rm \lg(M_{\odot}/L_{\odot})$ & Best-fit stellar mass-to-light ratio\\
  & log\_rho\_s  & $\rm \lg(M_{\odot}\ {\rm kpc}^{-3}$) & The characteristic density of NFW profile\\
  & rs  & kpc & The break radius of NFW profile\\
  & kappa &  & The ratio between modelled line-of-sight velocity field and the observed one\\
  & log\_Mt\_Re  & $\rm \lg(M_{\odot})$ & Enclosed total mass within a sphere of effective radius\\
  & log\_Ms\_Re  & $\rm \lg(M_{\odot})$ & Enclosed stellar mass within a sphere of effective radius\\
  & log\_Md\_Re  & $\rm \lg(M_{\odot})$ & Enclosed dark matter mass within a sphere of effective radius\\
  & fdm\_Re &  & Dark matter fraction within a sphere of effective radius\\
  & log\_ML\_dyn\_Re  & $\rm \lg(M_{\odot}/L_{\odot})$ & Dynamical mass-to-light ratio within effective radius\\
  & chi2\_dof &  & The reduced chi-square of the best-fit model (The values are scaled to \\&&&\ \ \ \ \ \ account for the effect of standard deviation of the $\rm \chi^2$ itself, \\&&&\ \ \ \ \ \ should be only used in the comparison between different models)\\
  & rhalf\_arcsec  & arcsec & Radius of the sphere which encloses half the total luminosity\\
  & log\_Mt\_rhalf  & $\rm \lg(M_{\odot}/L_{\odot})$ & Enclosed total mass within a sphere of 3D half-light radius\\
  & log\_Ms\_rhalf  & $\rm \lg(M_{\odot}/L_{\odot})$ & Enclosed stellar mass within a sphere of 3D half-light radius\\
  & log\_Md\_rhalf  & $\rm \lg(M_{\odot}/L_{\odot})$ & Enclosed dark matter mass within a sphere of 3D half-light radius\\
  & MW\_Gt\_Re &  & Mass-weighted total density slope within a sphere of effective radius\\
  & MW\_Gs\_Re &  & Mass-weighted stellar density slope within a sphere of effective radius\\
  & MW\_Gd\_Re &  & Mass-weighted dark matter density slope within a sphere of effective radius\\
  & MW\_Gt\_rhalf &  & Mass-weighted total density slope within a sphere of 3D half-light radius\\
  & MW\_Gs\_rhalf &  & Mass-weighted stellar density slope within a sphere of 3D half-light radius\\
  & MW\_Gd\_rhalf &  & Mass-weighted dark matter density slope within a sphere of 3D half-light radius\\
  & Gt\_Re &  & Average logarithmic total density slope between 0.1 and 1 effective radius\\
  & Gs\_Re &  & Average logarithmic stellar density slope between 0.1 and 1 effective radius\\
  & Gd\_Re &  & Average logarithmic dark matter density slope between 0.1 and 1 effective radius\\
 \hline
5 & inc\_deg &  & Best-fit inclination angle (being 90$^{\circ}$ for edge-on)\\
($\rm JAM_{sph}+$
  & beta\_r &  & Best-fit radial velocity anisotropy in spherical coordinates\\
NFW)  & log\_ML\_stellar  & $\rm \lg(M_{\odot}/L_{\odot})$ & Best-fit stellar mass-to-light ratio\\
  & log\_rho\_s  & $\rm \lg(M_{\odot}\ {\rm kpc}^{-3}$) & The characteristic density of NFW profile\\
  & rs &  kpc & The break radius of NFW profile\\
  & kappa &  & The ratio between modelled line-of-sight velocity field and the observed one\\
  & log\_Mt\_Re  & $\rm \lg(M_{\odot})$ & Enclosed total mass within a sphere of effective radius\\
  & log\_Ms\_Re  & $\rm \lg(M_{\odot})$ & Enclosed stellar mass within a sphere of effective radius\\
  & log\_Md\_Re  & $\rm \lg(M_{\odot})$ & Enclosed dark matter mass within a sphere of effective radius\\
  & fdm\_Re &  & Dark matter fraction within a sphere of effective radius\\
  & log\_ML\_dyn\_Re  & $\rm \lg(M_{\odot}/L_{\odot})$ & Dynamical mass-to-light ratio within effective radius\\
  & chi2\_dof &  & The reduced chi-square of the best-fit model (The values are scaled to \\&&&\ \ \ \ \ \ account for the effect of standard deviation of the $\rm \chi^2$ itself, \\&&&\ \ \ \ \ \ should be only used in the comparison between different models)\\
  & rhalf\_arcsec  & arcsec & Radius of the sphere which encloses half the total luminosity\\
  & log\_Mt\_rhalf  & $\rm \lg(M_{\odot}/L_{\odot})$ & Enclosed total mass within a sphere of 3D half-light radius\\
  & log\_Ms\_rhalf  & $\rm \lg(M_{\odot}/L_{\odot})$ & Enclosed stellar mass within a sphere of 3D half-light radius\\
  & log\_Md\_rhalf  & $\rm \lg(M_{\odot}/L_{\odot})$ & Enclosed dark matter mass within a sphere of 3D half-light radius\\
  & MW\_Gt\_Re &  & Mass-weighted total density slope within a sphere of effective radius\\
  & MW\_Gs\_Re &  & Mass-weighted stellar density slope within a sphere of effective radius\\
  & MW\_Gd\_Re &  & Mass-weighted dark matter density slope within a sphere of effective radius\\
  & MW\_Gt\_rhalf &  & Mass-weighted total density slope within a sphere of 3D half-light radius\\
  & MW\_Gs\_rhalf &  & Mass-weighted stellar density slope within a sphere of 3D half-light radius\\
  & MW\_Gd\_rhalf &  & Mass-weighted dark matter density slope within a sphere of 3D half-light radius\\
  & Gt\_Re &  & Average logarithmic total density slope between 0.1 and 1 effective radius\\
  & Gs\_Re &  & Average logarithmic stellar density slope between 0.1 and 1 effective radius\\
  & Gd\_Re &  & Average logarithmic dark matter density slope between 0.1 and 1 effective radius\\
  \hline
6 & inc\_deg &  & Best-fit inclination angle (being 90$^{\circ}$ for edge-on)\\
($\rm JAM_{cyl}+$
  & beta\_z &  & Best-fit radial velocity anisotropy in cylindrical coordinates\\
fixed NFW)  & log\_ML\_stellar  & $\rm \lg(M_{\odot}/L_{\odot})$ & Best-fit stellar mass-to-light ratio\\
  & log\_rho\_s  & $\rm \lg(M_{\odot}\ {\rm kpc}^{-3}$) & The characteristic density of NFW profile\\
  & rs  & kpc & The break radius of NFW profile\\
  & kappa &  & The ratio between modelled line-of-sight velocity field and the observed one\\
  & log\_Mt\_Re  & $\rm \lg(M_{\odot})$ & Enclosed total mass within a sphere of effective radius\\
  & log\_Ms\_Re  & $\rm \lg(M_{\odot})$ & Enclosed stellar mass within a sphere of effective radius\\
  & log\_Md\_Re  & $\rm \lg(M_{\odot})$ & Enclosed dark matter mass within a sphere of effective radius\\
  & fdm\_Re &  & Dark matter fraction within a sphere of effective radius\\
  & log\_ML\_dyn\_Re  & $\rm \lg(M_{\odot}/L_{\odot})$ & Dynamical mass-to-light ratio within effective radius\\
  & chi2\_dof &  & The reduced chi-square of the best-fit model (The values are scaled to \\&&&\ \ \ \ \ \ account for the effect of standard deviation of the $\rm \chi^2$ itself, \\&&&\ \ \ \ \ \ should be only used in the comparison between different models)\\
  & rhalf\_arcsec  & arcsec & Radius of the sphere which encloses half the total luminosity\\
  & log\_Mt\_rhalf  & $\rm \lg(M_{\odot}/L_{\odot})$ & Enclosed total mass within a sphere of 3D half-light radius\\
  & log\_Ms\_rhalf  & $\rm \lg(M_{\odot}/L_{\odot})$ & Enclosed stellar mass within a sphere of 3D half-light radius\\
  & log\_Md\_rhalf  & $\rm \lg(M_{\odot}/L_{\odot})$ & Enclosed dark matter mass within a sphere of 3D half-light radius\\
  & MW\_Gt\_Re &  & Mass-weighted total density slope within a sphere of effective radius\\
  & MW\_Gs\_Re &  & Mass-weighted stellar density slope within a sphere of effective radius\\
  & MW\_Gd\_Re &  & Mass-weighted dark matter density slope within a sphere of effective radius\\
  & MW\_Gt\_rhalf &  & Mass-weighted total density slope within a sphere of 3D half-light radius\\
  & MW\_Gs\_rhalf &  & Mass-weighted stellar density slope within a sphere of 3D half-light radius\\
  & MW\_Gd\_rhalf &  & Mass-weighted dark matter density slope within a sphere of 3D half-light radius\\
  & Gt\_Re &  & Average logarithmic total density slope between 0.1 and 1 effective radius\\
  & Gs\_Re &  & Average logarithmic stellar density slope between 0.1 and 1 effective radius\\
  & Gd\_Re &  & Average logarithmic dark matter density slope between 0.1 and 1 effective radius\\
  \hline
7 & inc\_deg &  & Best-fit inclination angle (being 90$^{\circ}$ for edge-on)\\
($\rm JAM_{sph}$+
  & beta\_r &  & Best-fit radial velocity anisotropy in spherical coordinates\\
fixed NFW)  & log\_ML\_stellar  & $\rm \lg(M_{\odot}/L_{\odot})$ & Best-fit stellar mass-to-light ratio\\
  & log\_rho\_s  & $\rm \lg(M_{\odot}\ {\rm kpc}^{-3}$) & The characteristic density of NFW profile\\
  & rs  & kpc & The break radius of NFW profile\\
  & kappa &  & The ratio between modelled line-of-sight velocity field and the observed one\\
  & log\_Mt\_Re  & $\rm \lg(M_{\odot})$ & Enclosed total mass within a sphere of effective radius\\
  & log\_Ms\_Re  & $\rm \lg(M_{\odot})$ & Enclosed stellar mass within a sphere of effective radius\\
  & log\_Md\_Re  & $\rm \lg(M_{\odot})$ & Enclosed dark matter mass within a sphere of effective radius\\
  & fdm\_Re &  & Dark matter fraction within a sphere of effective radius\\
  & log\_ML\_dyn\_Re  & $\rm \lg(M_{\odot}/L_{\odot})$ & Dynamical mass-to-light ratio within effective radius\\
  & chi2\_dof &  & The reduced chi-square of best-fit model (The values are scaled to \\&&&\ \ \ \ \ \ account for the effect of standard deviation of the $\rm \chi^2$ itself, \\&&&\ \ \ \ \ \ should be only used in the comparison between different models)\\
  & rhalf\_arcsec  & arcsec & Radius of the sphere which encloses half the total luminosity\\
  & log\_Mt\_rhalf  & $\rm \lg(M_{\odot}/L_{\odot})$ & Enclosed total mass within a sphere of 3D half-light radius\\
  & log\_Ms\_rhalf  & $\rm \lg(M_{\odot}/L_{\odot})$ & Enclosed stellar mass within a sphere of 3D half-light radius\\
  & log\_Md\_rhalf  & $\rm \lg(M_{\odot}/L_{\odot})$ & Enclosed dark matter mass within a sphere of 3D half-light radius\\
  & MW\_Gt\_Re &  & Mass-weighted total density slope within a sphere of effective radius\\
  & MW\_Gs\_Re &  & Mass-weighted stellar density slope within a sphere of effective radius\\
  & MW\_Gd\_Re &  & Mass-weighted dark matter density slope within a sphere of effective radius\\
  & MW\_Gt\_rhalf &  & Mass-weighted total density slope within a sphere of 3D half-light radius\\
  & MW\_Gs\_rhalf &  & Mass-weighted stellar density slope within a sphere of 3D half-light radius\\
  & MW\_Gd\_rhalf &  & Mass-weighted dark matter density slope within a sphere of 3D half-light radius\\
  & Gt\_Re &  & Average logarithmic total density slope between 0.1 and 1 effective radius\\
  & Gs\_Re &  & Average logarithmic stellar density slope between 0.1 and 1 effective radius\\
  & Gd\_Re &  & Average logarithmic dark matter density slope between 0.1 and 1 effective radius\\
  \hline
8 & inc\_deg &  & Best-fit inclination angle (being 90$^{\circ}$ for edge-on)\\
($\rm JAM_{cyl}+$
  & beta\_z &  & Best-fit radial velocity anisotropy in cylindrical coordinates\\
gNFW)  & log\_ML\_stellar  & $\rm \lg(M_{\odot}/L_{\odot})$ & Best-fit stellar mass-to-light ratio\\
  & log\_rho\_s  & $\rm \lg(M_{\odot}\ {\rm kpc}^{-3}$) & The characteristic density of gNFW profile\\
  & rs  & kpc & The break radius of gNFW profile\\
  & gamma\_gNFW &  & The inner density slope of gNFW profile\\
  & kappa &  & The ratio between modelled line-of-sight velocity field and the observed one\\
  & log\_Mt\_Re  & $\rm \lg(M_{\odot})$ & Enclosed total mass within a sphere of effective radius\\
  & log\_Ms\_Re  & $\rm \lg(M_{\odot})$ & Enclosed stellar mass within a sphere of effective radius\\
  & log\_Md\_Re  & $\rm \lg(M_{\odot})$ & Enclosed dark matter mass within a sphere of effective radius\\
  & fdm\_Re &  & Dark matter fraction within a sphere of effective radius\\
  & log\_ML\_dyn\_Re  & $\rm \lg(M_{\odot}/L_{\odot})$ & Dynamical mass-to-light ratio within effective radius\\
  & chi2\_dof &  & The reduced chi-square of the best-fit model (The values are scaled to \\&&&\ \ \ \ \ \ account for the effect of standard deviation of the $\rm \chi^2$ itself, \\&&&\ \ \ \ \ \ should be only used in the comparison between different models)\\
  & rhalf\_arcsec  & arcsec & Radius of the sphere which encloses half the total luminosity\\
  & log\_Mt\_rhalf  & $\rm \lg(M_{\odot}/L_{\odot})$ & Enclosed total mass within a sphere of 3D half-light radius\\
  & log\_Ms\_rhalf  & $\rm \lg(M_{\odot}/L_{\odot})$ & Enclosed stellar mass within a sphere of 3D half-light radius\\
  & log\_Md\_rhalf  & $\rm \lg(M_{\odot}/L_{\odot})$ & Enclosed dark matter mass within a sphere of 3D half-light radius\\
  & MW\_Gt\_Re &  & Mass-weighted total density slope within a sphere of effective radius\\
  & MW\_Gs\_Re &  & Mass-weighted stellar density slope within a sphere of effective radius\\
  & MW\_Gd\_Re &  & Mass-weighted dark matter density slope within a sphere of effective radius\\
  & MW\_Gt\_rhalf &  & Mass-weighted total density slope within a sphere of 3D half-light radius\\
  & MW\_Gs\_rhalf &  & Mass-weighted stellar density slope within a sphere of 3D half-light radius\\
  & MW\_Gd\_rhalf &  & Mass-weighted dark matter density slope within a sphere of 3D half-light radius\\
  & Gt\_Re &  & Average logarithmic total density slope between 0.1 and 1 effective radius\\
  & Gs\_Re &  & Average logarithmic stellar density slope between 0.1 and 1 effective radius\\
  & Gd\_Re &  & Average logarithmic dark matter density slope between 0.1 and 1 effective radius\\
  \hline
9 & inc\_deg &  & Best-fit inclination angle (being 90$^{\circ}$ for edge-on)\\
($\rm JAM_{sph}+$
  & beta\_r &  & Best-fit radial velocity anisotropy in spherical coordinates\\
gNFW)  & log\_ML\_stellar  & $\rm \lg(M_{\odot}/L_{\odot})$ & Best-fit stellar mass-to-light ratio\\
  & log\_rho\_s  & $\rm \lg(M_{\odot}\ {\rm kpc}^{-3}$) & The characteristic density of gNFW profile\\
  & rs  & kpc & The break radius of gNFW profile\\
  & gamma\_gNFW &  & The inner density slope of gNFW profile\\
  & kappa &  & The ratio between modelled line-of-sight velocity field and the observed one\\
  & log\_Mt\_Re  & $\rm \lg(M_{\odot})$ & Enclosed total mass within a sphere of effective radius\\
  & log\_Ms\_Re  & $\rm \lg(M_{\odot})$ & Enclosed stellar mass within a sphere of effective radius\\
  & log\_Md\_Re  & $\rm \lg(M_{\odot})$ & Enclosed dark matter mass within a sphere of effective radius\\
  & fdm\_Re &  & Dark matter fraction within a sphere of effective radius\\
  & log\_ML\_dyn\_Re  & $\rm \lg(M_{\odot}/L_{\odot})$ & Dynamical mass-to-light ratio within effective radius\\
  & chi2\_dof  & & The reduced chi-square of the best-fit model (The values are scaled to \\&&&\ \ \ \ \ \  account for the effect of standard deviation of the $\rm \chi^2$ itself, \\&&&\ \ \ \ \ \ should be only used in the comparison between different models)\\
  & rhalf\_arcsec  & arcsec & Radius of the sphere which encloses half the total luminosity\\
  & log\_Mt\_rhalf  & $\rm \lg(M_{\odot}/L_{\odot})$ & Enclosed total mass within a sphere of 3D half-light radius\\
  & log\_Ms\_rhalf  & $\rm \lg(M_{\odot}/L_{\odot})$ & Enclosed stellar mass within a sphere of 3D half-light radius\\
  & log\_Md\_rhalf  & $\rm \lg(M_{\odot}/L_{\odot})$ & Enclosed dark matter mass within a sphere of 3D half-light radius\\
  & MW\_Gt\_Re &  & Mass-weighted total density slope within a sphere of effective radius\\
  & MW\_Gs\_Re &  & Mass-weighted stellar density slope within a sphere of effective radius\\
  & MW\_Gd\_Re &  & Mass-weighted dark matter density slope within a sphere of effective radius\\
  & MW\_Gt\_rhalf &  & Mass-weighted total density slope within a sphere of 3D half-light radius\\
  & MW\_Gs\_rhalf &  & Mass-weighted stellar density slope within a sphere of 3D half-light radius\\
  & MW\_Gd\_rhalf &  & Mass-weighted dark matter density slope within a sphere of 3D half-light radius\\
  & Gt\_Re &  & Average logarithmic total density slope between 0.1 and 1 effective radius\\
  & Gs\_Re &  & Average logarithmic stellar density slope between 0.1 and 1 effective radius\\
  & Gd\_Re &  & Average logarithmic dark matter density slope between 0.1 and 1 effective radius\\
  \hline
\end{longtable}


\bsp	
\label{lastpage}
\end{document}